\documentclass[epj]{svjour}

\usepackage{graphicx}
\usepackage{graphics}
\usepackage{amssymb}
\usepackage{amsmath}
\usepackage{bm}
\usepackage{slashed}

\newcommand{\be}{\begin{equation}}
\newcommand{\ee}{\end{equation}}
\newcommand{\bea}{\begin{eqnarray}}
\newcommand{\eea}{\end{eqnarray}}
\newcommand{\beal}{\begin{align}}
\newcommand{\eal}{\end{align}}
\newcommand{\bespl}{\begin{split}}
\newcommand{\espl}{\end{split}}

\newcommand{\nslash}{\kern 0.2 em n\kern -0.50em /}
\newcommand{\kslash}{\kern 0.2 em k\kern -0.45em /}
\newcommand{\pslash}{\kern 0.2 em p\kern -0.50em /}
\newcommand{\Sslash}{\kern 0.2 em S\kern -0.50em /}
\newcommand{\Pslash}{\kern 0.2 em P\kern -0.50em /}
\newcommand{\Rslash}{\kern 0.2 em R\kern -0.50em /}

\begin{document}

\title{Effects 
of azimuth-symmetric acceptance cutoffs on the 
measured asymmetry in unpolarized Drell-Yan fixed target 
experiments} 
\subtitle{Acceptance cutoffs in Drell-Yan}

\author{A.~Bianconi\inst{1}  
\and M.P.~Bussa\inst{2} 
\and M.~Destefanis\inst{2} \and L.~Ferrero\inst{2} 
\and M.~Greco\inst{2} \and M.~Maggiora\inst{2} \and S.~Spataro\inst{2}}

\institute{Dipartimento di Ingegneria dell'Informazione -  
Universit\`a degli Studi di Brescia, and 
Istituto Nazionale di Fisica Nucleare, Gruppo di Brescia, 
via Valotti 9,  
I-25123 Brescia, Italy\\ \email{bianconi@bs.infn.it} 
\and 
Dipartimento di Fisica - Universit\`a degli Studi di Torino, Italy, 
and 
Istituto Nazionale di Fisica Nucleare, Sezione di Torino,  
via Giuria 1, I-10125 Torino, Italy}

\date{}

\abstract{
Fixed-target unpolarized 
Drell-Yan experiments often feature an acceptance depending 
on the polar angle of the lepton tracks in the laboratory frame. 
Typically leptons are detected in a defined angular range, 
with a dead zone in the forward region. 
If the cutoffs imposed by the angular acceptance  are independent 
of the azimuth, 
at first sight they do not appear dangerous for 
a measurement of the $cos(2\phi)$-asymmetry, relevant because of its   
association with the violation 
of the Lam-Tung rule and with the Boer-Mulders function. 
On the contrary, direct simulations show that up to 10 percent 
asymmetries are produced by these cutoffs. These artificial  
asymmetries present qualitative features that allow them to 
mimic the physical ones. 
They introduce some model-dependence in the measurements 
of the $cos(2\phi)$-asymmetry, since a precise reconstruction of the 
acceptance in the Collins-Soper frame requires a Monte Carlo simulation, 
that in turn requires some detailed physical input to generate event 
distributions. 
Although experiments in the eighties seem to have been 
aware of this problem, the possibility of using the Boer-Mulders 
function as an input parameter in the extraction of Transversity has 
much increased the requirements of precision on this measurement. 
Our simulations show that the safest approach to these measurements 
is a strong cutoff on the 
Collins-Soper polar angle. This reduces statistics, but does not 
necessarily decrease the precision in a measurement of the 
Boer-Mulders function. 
}

\PACS{ 
{13.85.Qk} {Drell-Yan} \and
{13.88.+e} {Polarization in interaction and scattering} \and
{29.30.-h} {Spectrometers and spectroscopic techniques}
}

\maketitle

\section{Introduction}

\subsection{The physics case}

Drell-Yan experiments\cite{DrellYan70}  have a long history (a brief 
introduction may 
be found in chapter 5 of \cite{Field} and in the 
reviews\cite{Matthiae} and 
\cite{Kenyon82}, while an extensive data collection has been 
organized in the Hepdata database\cite{HEPDATA}. A more modern 
general scheme may be found in \cite{metz}). 

One of the most interesting and controversial observables measured in 
unpolarized Drell-Yan  
is the so-called $\nu$ coefficient\cite{LamTung78a,LamTung78b} 
associated with the 
$cos(2\phi)$ azimuthal asymmetry in the rest frame of the dilepton pair. 
This  
quantity has theoretical relevance because of its association with 
the violation of the PQCD Lam-Tung 
relation\cite{LamTung78a,LamTung78b}, and with the T-odd Boer-Mulders 
distribution function\cite{BoerMulders98,Boer99}. 

Recently, its relevance has been increased by the 
perspective of using it to extract transversity\cite{RalstonSoper79} 
from the combination of single-spin and unpolarized Drell-Yan 
measurements\cite{Boer99}. This however requires the $\nu$ coefficient 
to be measured at an unprecedented level of precision: a result like 
$10 \pm 3$\% in a measurement of the $cos(2\phi)$ asymmetry would 
be interesting in itself, but useless for a quantitative identification 
of transversity that may compete with other techniques. 

This work is centered on the potential errors   
that a beam-related forward dead cone in the acceptance 
may introduce into a 
measurement of the $cos(2\phi)-$asymmetry. The subtle point is that 
an acceptance that does not depend on the azimuth in the 
laboratory frame is able to introduce azimuthal effects 
in the Collins-Soper frame\cite{CollinsSoper77} (or similar frames) 
where this asymmetry is measured.  
Distributions of the acceptance that depended on the azimuthal angle 
in the Collins-Soper frame were reported 
in works\cite{NA3a,NA3b,NA10a,NA10b,Conway89} presenting measurements of the 
$cos(2\phi)$ asymmetry in the years 1980-90. 
Here we use Monte Carlo simulations to 
analyze in detail and extensively 
the effects of an azimuth-symmetric forward dead cone on the measured 
$cos(2\phi)$-asymmetry. We explore a range of beam energies going from 
15 to 250 GeV, that is of interest for some proposals 
aimed at measuring the $cos(2\phi)$ asymmetry with high precision 
in the 
near future\cite{panda,pandapb,pax,maggiora1,compassDY}.

\subsection{General definitions}

In this work we have used the following definitions: 

\begin{itemize}

\item Laboratory frame: it is the frame where the target is at rest, 
with $z$ axis along the beam direction. 

\item $s$ is the squared center of mass energy per target nucleon.   
It is expressed in GeV$^2$/nucleon, and 
calculated as if the target 
consisted of one nucleon only. 

\item $x_1$ and $x_2$ are the longitudinal fractions of the 
annihilating partons: $x_1$ is from the projectile, and $x_2$ 
from the target (explicitly indicated as $x_{target}$ in the figures). 

\item $q_\mu$ $=$ $(E_\gamma,\vec q)$ 
is the 4-momentum of the virtual photon in the laboratory 
frame. 

\item $Q_T$ $=$ $\sqrt{q_x^2+q_y^2}$ 
is the modulus of the transverse momentum  
of the virtual photon w.r.t. the beam axis. 

\item $Q$ $=$ $\sqrt{Q^2}$ is the ``mass'' of the virtual photon, or 
equivalently the invariant mass of the lepton pair, and in the following 
will be named ``mass''. $Q$ is completely determined by $s$, $x_1$, 
$x_2$, $Q_T$. A standard approximation is $Q$ $\approx$ $x_1x_2s$, 
but this approximation gives the mass of the 4-vector $(Q_0,0,0,Q_z)$
and fails if $Q_T$ is not much smaller than $x_1x_2s$, that is the rule 
in some of the below examined kinematics. 

\item $\theta$ and $\phi$ are the polar and azimuthal angles 
of $one$ 
of the two leptons (since now on: the positive one) in a 
frame where the virtual photon is at rest and the leptons have 
opposite momenta. Although different 
choices exist for such a frame,
here we will use 
the Collins-Soper frame\cite{CollinsSoper77}. 

\item $\vec q^+$ and $\vec q^-$ are the lepton momenta in the laboratory  
frame, with $\vec q^+$ $+$ $\vec q^-$ $=$ $\vec q$. 

\item $\theta_{Lab+}$ and $\theta_{Lab-}$ are the angles of the positive 
and negative muon in the laboratory frame, where the target is at 
rest and the $z$-axis is parallel to the momentum $\vec P_{beam}$. 
No term in the cross section depends on these angles, but in this work 
they are central because acceptance cutoffs are imposed on them. 

\item Collision plane: it is the plane, in the laboratory frame, 
containing $\vec q$ and the beam axis. 

\item Lepton plane: it is the plane, in the laboratory frame, 
containing the two lepton momenta (and $\vec q$). 

\end{itemize}

The cross section for Drell-Yan dilepton production in scattering of 
unpolarized hadrons (charged pion, proton or antiproton vs a target 
nucleus) may be approximately written in the parton model form (see e.g. 
\cite{Field}): 
\begin{equation}
{d\sigma\over {dx_1 dx_2 dQ_T d\Omega}} \ =\ 
{1 \over s}\cdot 
W(x_1,x_2,Q_T)\cdot A(\theta, \phi).
\label{eq:e1}
\end{equation}
In 
unpolarized Drell-Yan, and far from kinematic regions where the virtual photon 
is dominated by quarkonium resonances,   
$A$ has the 
form\cite{LamTung78a}: 
\begin{align}
&A(\theta, \phi) \ =\ 
1\ +\ cos^2(\theta)\ +\ \nonumber\\ 
&+\ 
{{\nu(s,x_1,x_2,Q_T)} \over 2} sin^2(\theta)cos(2\phi)\  
+\ ....
\label{eq:e2}
\end{align}

$W(x_1,x_2,Q_T)$ does not depend on the angles. 
$A(\theta,\phi)$ in principle depends on $s$, $x_1$, 
$x_2$, $Q_T$ too. 

The focus of the present work is on the $\nu cos(2\phi)$ 
term\cite{LamTung78a} present in $A(\theta,\phi)$. The idea is that 
an acceptance constraint $\theta_{Lab\pm}$ $>$ $\theta_{cutoff}$ 
produces an unphysical contribution to the $\nu cos(2\phi)$-term. 

\subsection{Collins-Soper frame and $cos(2\phi)$-asymmetry}

Three relevant coordinate frames will be here considered: (1) The Laboratory 
frame, where 
the target is at rest, (2) the ``Collider'' frame, i.e. 
the center of mass frame of the projectile 
and of the hit nucleon, (3) the Collins-Soper frame. 

The angular cutoff that is relevant in this work is imposed on the 
polar angle of each lepton in the Laboratory frame. 
The Collider frame is necessary as an intermediate step for 
calculating the longitudinal fractions and some auxiliary variables 
needed to define the Collins-Soper frame (for each event, 
we have a different Collins-Soper frame). 
The difference $\vec p_1-\vec p_2$ of the 3-momenta 
of the colliding hadrons in the Collider frame identifies a direction, 
that is chosen as the 
$z$ axis of the Collins-Soper frame. The momentum $\vec q^*$
of the 
virtual photon (of the dilepton pair) in the collider frame identifies 
the $xz$ plane in the Collins-Soper frame. The angles $\theta$ 
and $\phi$ are the polar 
and the azimuthal angles of the positive lepton in this frame.  

Some intuitive qualitative features of the Collins-Soper frame 
and of the $cos(2\phi)$ asymmetry are reported in 
the Discussion section. Both the lepton 
plane and the collision planes contain $\vec q$, 
and in most of the events (not in all) the angle that in the 
laboratory frame 
expresses the relative orientation 
of the lepton and collision planes around $\vec q$ is approximately 
the Collins-Soper azimuthal angle $\phi$.  
In events  with positive $cos(2\phi)$ the two planes are roughly 
parallel, as for the lepton pair $A^+A^-$ in fig.\ref{fig:G2}. 
In events with negative $cos(2\phi)$ the two planes are roughly 
perpendicular, as for the lepton pair $B^+B^-$ in the same figure. 
There is no sensitivity to dipolar (lepton exchange) 
effects, since $cos(2\phi)$ $\rightarrow$ 
$cos(2\phi+2\pi)$ when the two leptons are exchanged. 

\subsection{Acceptance on single muons, and on muon pairs}

In a fixed target Drell-Yan experiment it is common to 
have a dead forward cone in the laboratory, and often a dead 
backward cone, where lepton tracks are invisible or submerged by 
noise. In other words, it is quite normal to have the acceptance limits 
$\theta_{forward}$ $<$ 
$\theta_{Lab\pm}$ $<$ 
$\theta_{backward}$. 
In the forward direction, the cone occupied by the beam and by the 
diffraction products of the hadron collisions is a zero acceptance 
region. For a beam energy of magnitude 100 GeV this normally means 
a few degrees, decreasing at increasing beam energy. 

Although both a forward and a backward cutoff produce effects 
like the ones discussed in the 
following, in this work we will limit the discussion to the 
effects of a forward cutoff. 
We will also restrict to the $s$ range 30-500 GeV$^2$.
This region includes the available 
measurements\cite{NA3a,NA3b,NA10a,NA10b,Conway89} of a nonzero 
value for 
$\nu$\footnote{More 
recently $\nu$ has 
been found compatible with zero by E866\cite{E866}, but 
in a high-energy/small-$x$ regime that makes this measurement 
peculiar. We will not care such situation 
here.}, and 
some proposals for measuring it in the 
near future\cite{panda,pandapb,pax,maggiora1,compassDY}.

\begin{figure}[ht]
\centering
\includegraphics[width=9cm]{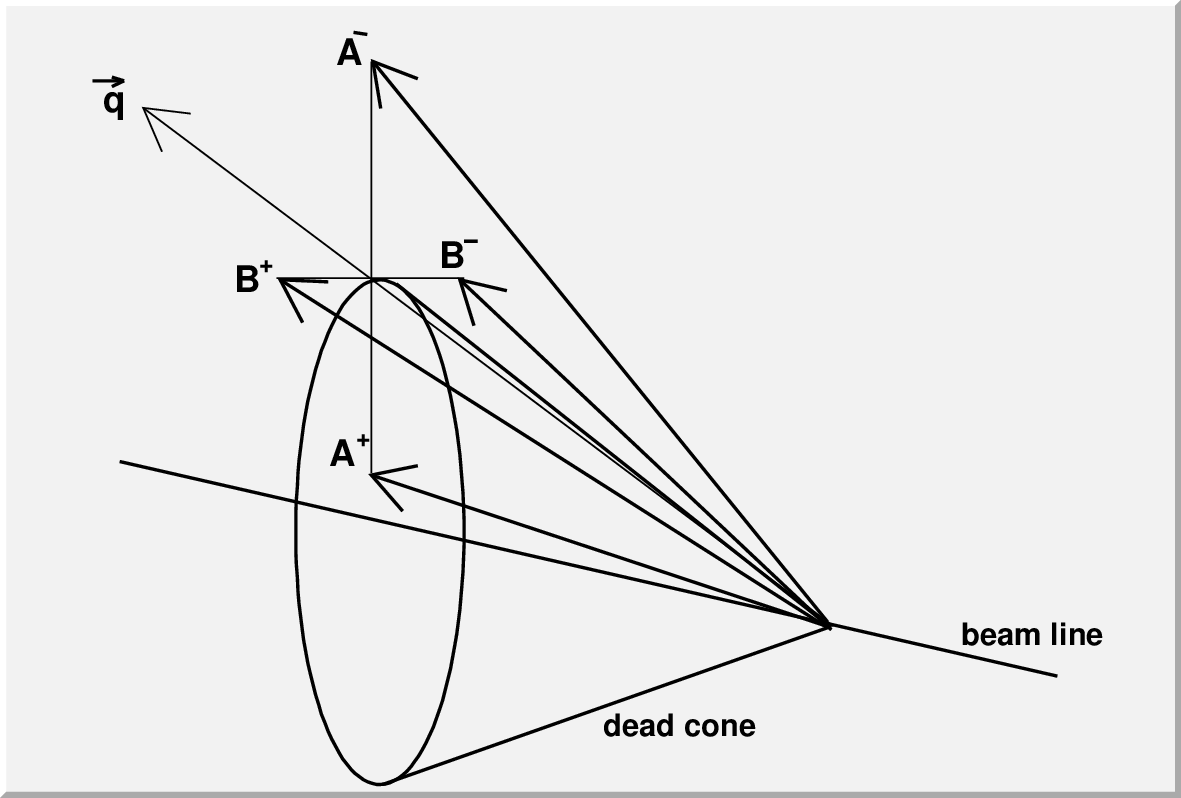}
\caption{
Schematic representation of the described effect. 
$A^+$ and $A^-$ are the tracks of a lepton pair composing an 
event where the lepton plane and the collision plane are parallel, 
while $B^+$ and $B^-$ are the tracks of a lepton pair in an 
event where the lepton plane and the collision plane are 
perpendicular. For both events $\vec q$ $=$ $\vec q^+$ $+$ 
$\vec q^-$ is the same and is tangent to the forward dead cone. 
Evidently, one of the four lepton tracks falls inside 
the dead cone ($A^+$), so that the $A^+A^-$ event is not 
detected. The $B^+B^-$ event 
is detected. 
\label{fig:G2}}
\end{figure}

If the effect of a forward dead cone on single lepton tracks is 
evident, much more subtle is the effect it may have on lepton $pairs$. 
The key qualitative statement of this work is that 
a forward dead cone removes preferentially events 
where the lepton plane and the collision plane are parallel. This may 
be seen in the peculiar example shown in fig. \ref{fig:G2}. 

The relevant elements appearing in this figure are: 
 
1) a forward dead cone (with enlarged  opening with respect to 
reality, for illustration purposes);

2) two possible 
lepton pairs ($A^+A^-$ and $B^+B^-$) with the same 
$\vec q$ $=$ $\vec q^+$ $+$ $\vec q^-$, 
but different space orientation for the lepton-antilepton plane;

3) $\vec q$ is chosen so 
to be tangent to the dead cone surface. 

In the $A^+A^-$ case one of the two leptons 
falls inside the dead cone. Since the pair is accepted only if two 
lepton tracks with opposite charge are detected, this pair is removed 
from the event collection of the experiment. 
In the $B^+B^-$ case neither $B^+$ nor $B^-$ is 
inside the dead cone, so this pair is detected and enters the 
event collection of the experiment. 
The $A^+A^-$ pair has negative $cos(2\phi)$, the $B^+B^-$ pair 
positive $cos(2\phi)$. All the other variables are equal for these two 
pairs. Therefore, by removing the former pair a false negative 
asymmetry is created.

Generalizing this example, a forward dead cone 
introduces a systematic anisotropy with respect to $cos(2\phi)$ 
into the distribution of the detected dilepton pairs. 
With suitable Monte Carlo simulations it is possible to study 
in a systematic way how large this effect can be. 

\subsection{Plan of this work}

Section II is mainly devoted to the ways we have 
selected and organized the simulated data. 
Section III is developed around 
figures showing the simulation results: some examples of 
possible experiment output, dependence of the effects on the most 
important variables, interplay between physical and artificial 
effects. Section IV presents a discussion about the 
obtained results and the more general problems raised by such 
analysis. 

\section{Methods}

\subsection{Relevant kinematics}

Two kinematic setups have special relevance in the following: 

\noindent
-) ``{\bf PANDA configuration}'': antiproton projectile on a nuclear target  
($Z/A$$=$ 0.4), with 
\begin{itemize}
	\item the squared c.m. energy is $s$ $=$ 30 GeV$^2$/nucleon;
	\item the dilepton/virtual photon invariant mass $Q$  satisfies 
1.8 GeV $<$ $Q$ $<$ 2.7 GeV;
	\item the pair overall transverse momentum $Q_T$ satisfies 
0.5 GeV/c $<$ $Q_T$ $<$ 2 GeV/c. 
\end{itemize}
In this configuration, inspired to 
the Drell-Yan program of the PANDA 
experiment\cite{panda,pandapb}, the relevant forward cutoff 
angles range from zero to about 10 degrees. 

\noindent
-) ``{\bf High-energy configuration}'': negative pion projectile on a nuclear 
target ($Z/A$ $=$ 0.4), with 
\begin{itemize}
	\item the squared c.m. energy is $s$ $=$ 500 GeV$^2$/nucleon;
	\item the dilepton/virtual photon invariant mass $Q$  satisfies 
4 GeV $<$ $Q$ $<$ 8 GeV;
	\item the pair overall transverse momentum $Q_T$ satisfies 
1 GeV/c $<$ $Q_T$ $<$ 2.5 GeV/c. 
\end{itemize}
In this configuration, inspired by 
past\footnote{ 
The high-energy choice is close to the conditions 
of the experiments E615\cite{Conway89} ($s$ $=$ 474 GeV$^2$/nucleon) 
and NA10\cite{NA10a,NA10b} running at its maximum beam energy   
($s$ $=$ 537 GeV$^2$/nucleon). 
} 
and 
planned\footnote{ 
In the planned Drell-Yan at Compass-II\cite{compassDY} $s$ ranges 
from 200 to 400 GeV$^2$/nucleon. 
}
pion-nucleus experiments, the relevant forward 
cutoff angles range from zero to about 1 degree. 

We will mainly refer to events generated in the phase space associated to 
these two configurations, and to obvious modifications of them. 

\subsection{Asymmetry definitions}

The asymmetry is here defined as in \cite{BRMCa,BRMCb,BRMCc,BRMCd}. 
Let us assume that the phase space of the experiment is divided into 
bins associated to the values of variables like $x$, $Q$ 
$Q_T$, $\theta$, but excluding the azimuthal variable $\phi$. So, the 
bin $i$ contains events with $\phi$ spread over all the range 
$-180^o$ $<$ $\phi$ $<$ $180^o$. Let us divide these events into two classes 
and define the {\bf individual bin asymmetry}:

\begin{align}
&f_{i+}\ \leftrightarrow\ \cos(2\phi)\ >\ 0\ in\ bin\ i\\
&f_{i-}\ \leftrightarrow\ \cos(2\phi)\ <\ 0\ in\ bin\ i\\
&population_{(bin\ i)}\ =\ {f_{i+}\ +\ f_{i-}}\\ 
&{\bf asymmetry_{(bin\ i)}}\ \equiv\ 
{{f_{i+}\ -\ f_{i-}} \over {f_{i+}\ +\ f_{i-}}} 
\label{eq:asimmetry_bin1}
\end{align}

We also need the {\bf full phase space integrated asymmetry}. 
This expression means that the 
complete phase space of the simulation (e.g. the phase space of 
the above described PANDA configuration) is divided into two 
halves corresponding to positive and negative values of $cos(2\phi)$. 


Using this way for calculating the $\nu-$related 
asymmetry 
is especially simple, and in absence of any 
forward dead cone it 
would lead to measured 
asymmetries that are roughly $\approx$ $\nu/2$. 
As it is evident from eq.\ref{eq:e2}, 
$\nu$ coincides with 
$(f_+-f_-)/(f_++f_-)$ when this quantity is estimated from a subset 
of events with $\theta$ $\approx$ 90$^\circ$. To include the full phase 
space (or at least the region $|cos(\theta)|$ $<$ 0.8-0.9, as it 
was customary in past experiments) means to 
reduce the resulting asymmetry by about a factor 2, since 
$sin^2(\theta)$ reduces $f_+-f_-$ while the 
factor $1+cos^2(\theta)$ increases $f_++f_-$ in eq.\ref{eq:e2}.

\subsection{Physical asymmetries}

Some of our simulations refer to a hypothetical experiment where an 
azimuthal asymmetry is found and its origin is completely artificial. 
In other cases however, we will consider the combination of physical 
and artificial effects. 

We will consider the possibility of a physical $cos(2\phi)$ asymmetry 
whose full phase space integrated values range  
between approximately $-20$ \% and $+20$ \%, analyzing the combined effect of 
this physical asymmetry and of an angular cutoff. 

We also consider the effect of a physical $cos(\phi)$ asymmetry 
in affecting the relation between cutoff angles and false 
$cos(2\phi)$ asymmetries. We will not present the effects of the cutoffs 
on the measured $cos(\phi)$ asymmetry itself, because this would 
make the present work too much long. 

The values of physical asymmetries employed in this work 
vary in ranges whose limits are $\pm$ 13.5 \% for the  
$cos(2\phi)$ asymmetry, $\pm$ 16 \% for the $cos(\phi)$ asymmetry in 
PANDA configuration, $\pm$ 22 \% for the $cos(\phi)$ asymmetry in the 
high energy configuration. These values are ``borderline'' in the sense that 
with stronger asymmetries the Monte Carlo simulation meets 
negative value of the cross section somewhere at large $Q_T$. 


\subsection{Simulations: the generator code}

The Monte Carlo simulation of Drell-Yan pairs is performed with the  
generator code\cite{DY_AB5} used in \cite{BRMCa,BRMCb,BRMCc,BRMCd}. 
All details about the event generation technique may be found 
in these references. A critical discussion of the 
underlying formalism may be found in \cite{BR_JPG2}. 
The cross section used for generating the events has the form 
given in eqs. \ref{eq:e1} and \ref{eq:e2}, with the parameters tuned to 
those values that reproduce the results of the 
experiments \cite{NA3a,NA3b,NA10a,NA10b,Conway89} in the pion-nucleus case. 
For the antiproton-nucleus 
case the parameters of the cross section present a certain amount 
of modeling for lack of antiproton-nucleus data at the required 
center of mass energy. 

\subsection{Simulations: organization of the simulated events}

In absence of angular cutoffs, the standard simulation used in 
this work produces 300,000 events.  
After applying acceptance cutoffs, the number $N$ of events to be analyzed  
is in the range 120,000-300,000.  
These events have been treated in different ways: 

\begin{itemize}

\item {\bf Data Analysis Method 1}: The $N$ events are considered 
as six independent 
simulations of an experiment that 
collects $N/6$ events. In each experiment the $N/6$ events 
may be further distributed into bins. 
The six repetitions of the simulation are treated as six independent 
experiments, and 
are exploited to estimate (i) the asymmetry measured in each bin as the 
average among the six values measured in each experiment for that bin, 
(ii) the error from an analysis of the fluctuations of the six 
measurements. 

\item {\bf Data Analysis Method 2}: The simulated $N$ events form a unique set, 
and the error in the asymmetry 
is estimated theoretically as $1/\sqrt{N}$ assuming a binomial distribution 
for the population of each of the two subsets $a$ and $b$ that we use 
to calculate the asymmetry as $(a-b)/(a+b)$. 

\item {\bf Subset populations}: In some cases we have distributed the $N$  
events of a simulated  
sample into bins of some variable like $x$ or $Q_T$. This strategy leads 
to bin populations and error bars that reproduce those of a real measurement. 
In the case of $Q$-binning over a wide $Q$-range we have preferred to 
produce an independent simulation for each $Q$-range, each 
with $N$ events.  
This strategy was necessary 
to have reasonable error bars at large $Q$, without being obliged 
to produce huge numbers of events over the entire phase space. 
The relative size of the error bars at different 
$Q$ does not reflect what would happen in a real experiment. 
\end{itemize}

Method 1 is the one used in \cite{BRMCa}, where 50,000 events 
were identified as the minimum for a meaningful 
measurement of the relevant features of the $cos(2\phi)$ asymmetry as 
a function of $x$ and $Q_T$, in a $Q$ range of width 1-2 GeV above the 
applied threshold. 

It was tested\cite{BRMCa} that the average values  
and the errors were stable enough if the number of 
repetitions was increased over six. However, 
the error estimated by this method is a random 
variable itself, and this is evident in some 
figures where nearby bins with very similar populations present 
different error bars. 

Two kinds of error may be extracted 
this way. 
An average asymmetry calculated 
with Method 1 is the average 
of 6 asymmetries $a_1$, $a_2$, $a_3$, $a_4$, $a_5$, $a_6$: 
$a_{final}$ $=$ $(\sum a_i)/6$. 
Each $a_i$ is extracted from $N/6$ events, 
so that $a_{final}$ is approximately the asymmetry extracted from a set 
of $N$ events\footnote{
``Approximately'' because an asymmetry $(a-b)/(a+b)$ 
is not linear in $a$ and $b$, so the 
average of 6 asymmetries, each from 
50,000 events, is not exactly the asymmetry from 300,000 
events. For small asymmetries the approximation is 
precise. 
}.

The 1-$\sigma$-error $\delta$ extracted as 
$\delta^2$ $=$ $\sum (a_i-a_{final})^2/5$ is an estimator 
for the error on $one$ of the six values $a_1$, $a_2$, ....$a_6$. 
We will name this error ``{\bf individual experiment error}''. 
On the other side, the error on the average of six identical 
measurements is 
$\sqrt{6}$ times smaller than the individual error. When we use 
$\delta_{final}$ $=$ 
$\delta_{final}/\sqrt{6}$ as an estimator for the error we write 
``{\bf error on the theoretical estimate}''. 

In this work Method 2 has been applied to full phase space 
asymmetries, extracted from event samples  
with $N$ ranging from 300,000 (no cuts) to 120,000 (the most severe cuts 
applied in this work, see later). 
The corresponding 1-$\sigma$ error on the asymmetries is 0.002-0.003. 
This may be smaller 
than the errors of numeric origin (these may be estimated from 
irregularities in the presented curves). On the other side, 
the error in Method 1 is a combination 
of statistical and statistically distributed numeric errors, and for this 
reason is frequently larger than $1/\sqrt{N}$. 


\subsection{Forward angular cutoffs}
 
In the following, the expression 
``applying a cutoff angle $\theta_{cutoff}$'' means that this event is 
excluded from the analyzed sample unless the individual lepton angles 
in the laboratory satisfy: 
\begin{align}
\theta_{Lab+}\ &>\ \theta_{cutoff},\\
\theta_{Lab-}\ &>\ \theta_{cutoff}.
\end{align} 
In absence of further specifications, in the present work 
``cutoff'' means a cutoff applied 
to the polar angles in the laboratory frame.

\section{Results}

In all our simulations the fake asymmetry comes out to be 
negative. In the following, we have reproduced it as positive  
in all those figures where it appears alone 
and its sign has no role. When a fake and a true asymmetry 
are compared or mixed their relative sign has relevance. In these 
cases we have reproduced 
all the true and fake asymmetries with their proper 
sign. 

\subsection{Angular distribution of the events in the 
Collins-Soper frame}

\begin{figure}[ht]
\centering
\includegraphics[width=9cm]{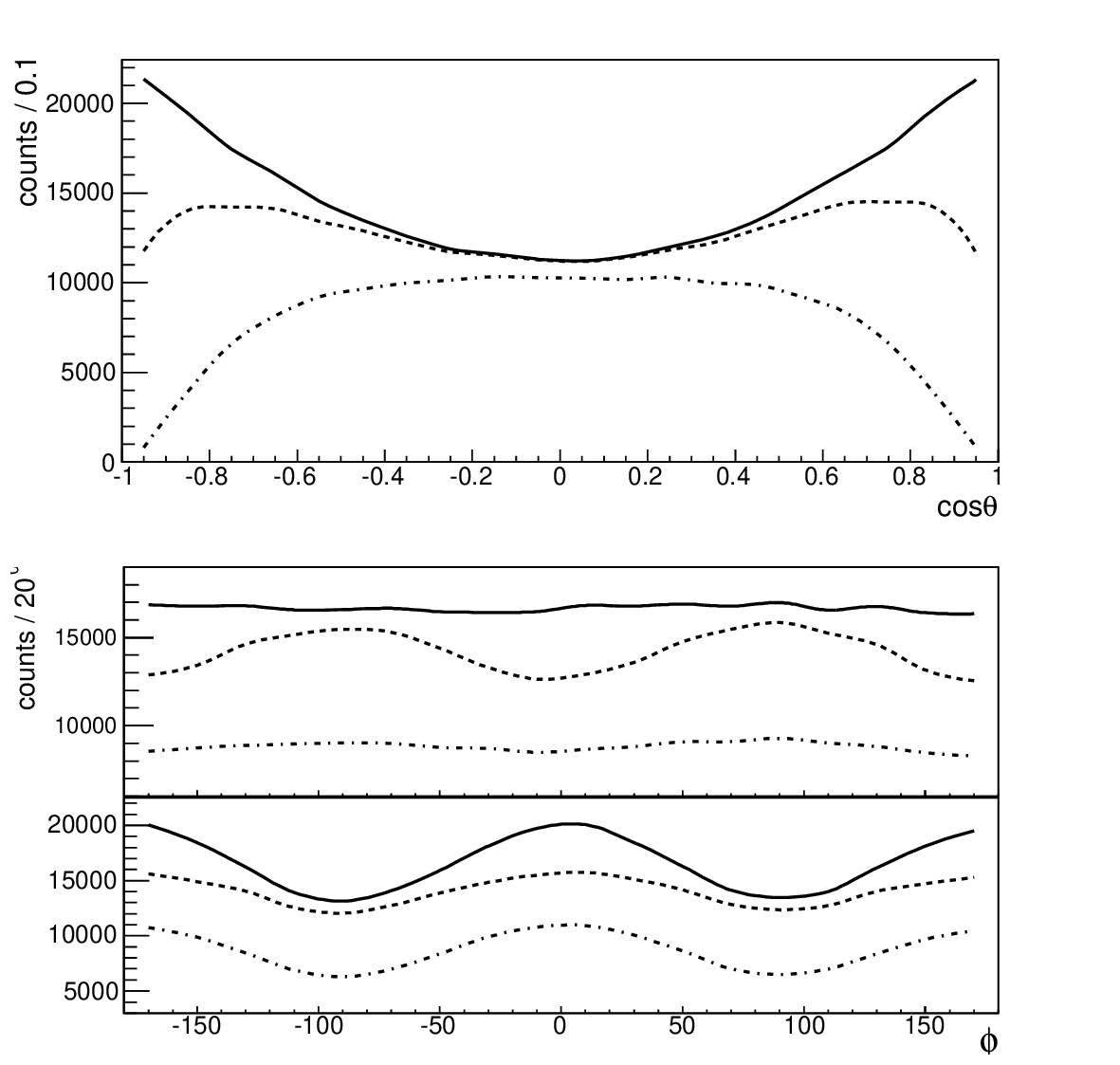}
\caption{
Angular distributions in the Collins-Soper frame in PANDA configuration. 
In each panel, the continuous line 
corresponds to no cutoff, the dashed line to cutoff angles 5$^o$ 
and the dash-dotted line to cutoff angle 10$^o$.   
Top panel: $cos(\theta)$ distributions, integrated w.r.t. $\phi$.
Bottom panel: $\phi$ distributions, integrated w.r.t. $\theta$. 
Upper part of the bottom panel: no physical asymmetry. 
Lower part: physical asymmetry with full phase space value 16 \%. 
\label{fig:ad1}}
\end{figure}

\begin{figure}[ht]
\centering
\includegraphics[width=9cm]{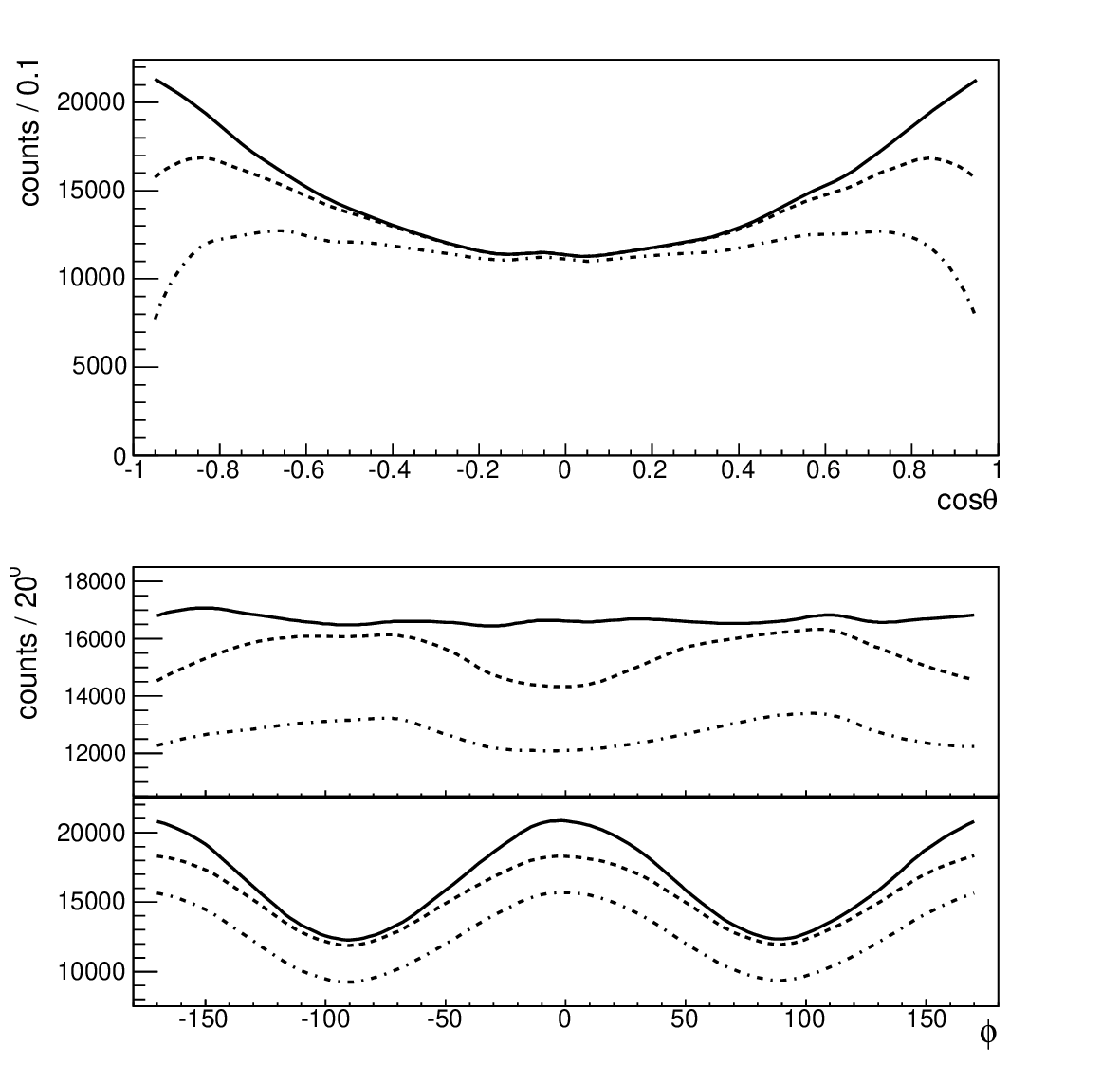}
\caption{
Angular distributions in the Collins-Soper frame in high energy 
configuration. 
In each panel, the continuous line 
corresponds to no cutoff, the dashed line to cutoff angles 0.5$^o$ 
and the dash-dotted line to cutoff angle 1$^o$.  
Top panel: $cos(\theta)$ distributions, integrated w.r.t. $\phi$.
Bottom panel: $\phi$ distributions, integrated w.r.t. $\theta$. 
Upper part of the bottom panel: no physical asymmetry. 
Lower part: physical asymmetry with full phase space value 22 \%. 
\label{fig:ad2}}
\end{figure}

The angular distributions of the events in the Collins-Soper frame 
in presence of physical asymmetries and/or angular cutoffs are 
presented in figures \ref{fig:ad1} and \ref{fig:ad2}.

Fig.\ref{fig:ad1} refers to PANDA configuration. 
In the upper panel 
we report the $cos(\theta)$-distribution of the events for  
cutoff angles $0^o$, $5^o$, $10^o$. 
In absence of acceptance cutoffs the $\phi$-integrated 
$\theta$-distribution must follow the $1+cos^2(\theta)$ law  
expected from eq.\ref{eq:e2}, since the other terms are 
removed by $\phi$-integration. Therefore, the presence of a $physical$ 
azimuthal asymmetry does not affect the $1+cos^2(\theta)$ 
shape. 
On the contrary, the same panel shows that a forward cutoff 
in the laboratory removes events from the regions $\theta$ $\approx$ 
0$^o$ and $\theta$ $\approx$ 180$^o$. 

The lower panel is divided into 
two parts. 
In the upper part, we report the $\phi$-distributions of the events 
for cutoff angles  $0^o$, $5^o$, $10^o$, in absence of any 
physical asymmetry effect. 
The no-cutoff curve is  
flat showing axial isotropy of the events in absence of a cutoff. 
For cutoff $5^o$, we have the maximum $cos(2\phi)$-effect, i.e. 
a distribution with two peaks within the $2 \pi$-range. We notice 
that this artificial $cos(2\phi)$-term is negative. For cutoff 10$^o$ 
the fake asymmetry is still present, but barely visible. 

The lower part of the lower panel shows the effect of the same angular 
cutoffs $0^o$, $5^o$, $10^o$ in presence of a strong physical 
asymmetry with positive sign 
(16 \% full-phase-space integrated asymmetry). The two effects 
seem to interfere in a linear way, i.e. the sum of a large and 
positive physical asymmetry and of a smaller negative artificial 
asymmetry is a small positive asymmetry. 
The curve corresponding to cutoff 10$^o$ is almost as flat as the 
no-cutoff curve, suggesting that at 10$^o$ the fake asymmetry 
contribution is negligible. 

The same analysis is reported in fig.\ref{fig:ad2} for the 
high-energy configuration. 
In this case the cutoff angles are 0$^o$, $0.5^o$, $1^o$, and the 
full phase space physical asymmetry is 22 \%. 
The qualitative conclusions are the same of the PANDA case. 

\subsection{Dependence of the fake asymmetry on the longitudinal 
fraction of the target}

\begin{figure}[ht]
\centering
\includegraphics[width=9cm]{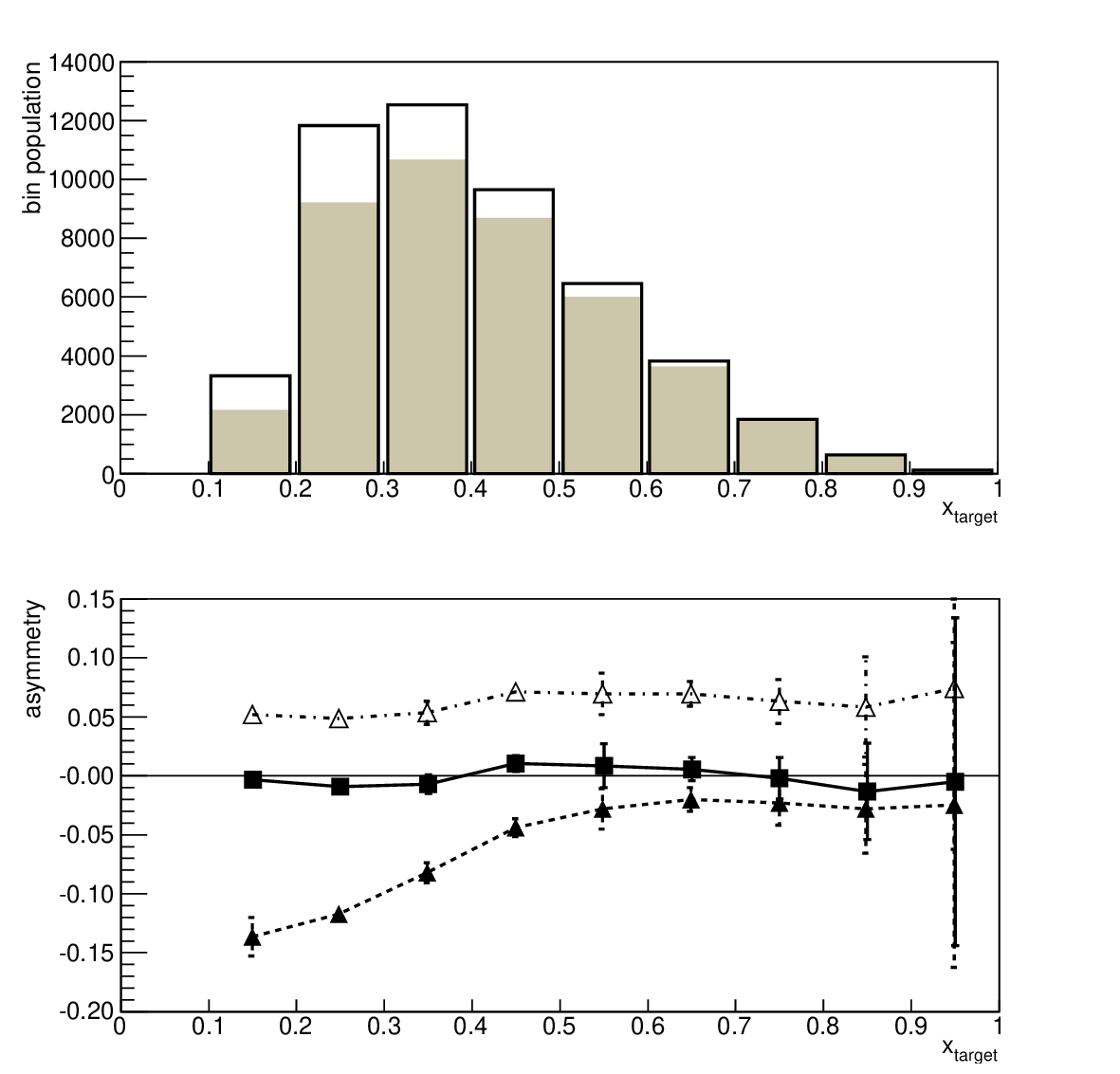}
\caption{
Simulated experiment in PANDA configuration with Method 1. 
In the upper plot the bin population is shown for each $x$ bin, without 
angular cutoff (black-line histogram) and with the cutoff 
$\theta_{lab\pm}$ $>$ 5$^\circ$ (shadowed histogram). 
In the lower plot, the extracted 
$cos(2\phi)$ asymmetry is shown. 
Continuous line: no physical 
asymmetry and no angular cutoff. 
Dot-dashed line: 
physical asymmetry 
with full phase space value 6.5 \%, 
no angular cuts. Dashed line: no physical asymmetry, 
cutoff $\theta_{lab\pm}$ $>$ 5$^\circ$ cut.
Error bars represent individual experiment errors. 
\label{fig:Xpanda}}
\end{figure}

\begin{figure}[ht]
\centering
\includegraphics[width=9cm]{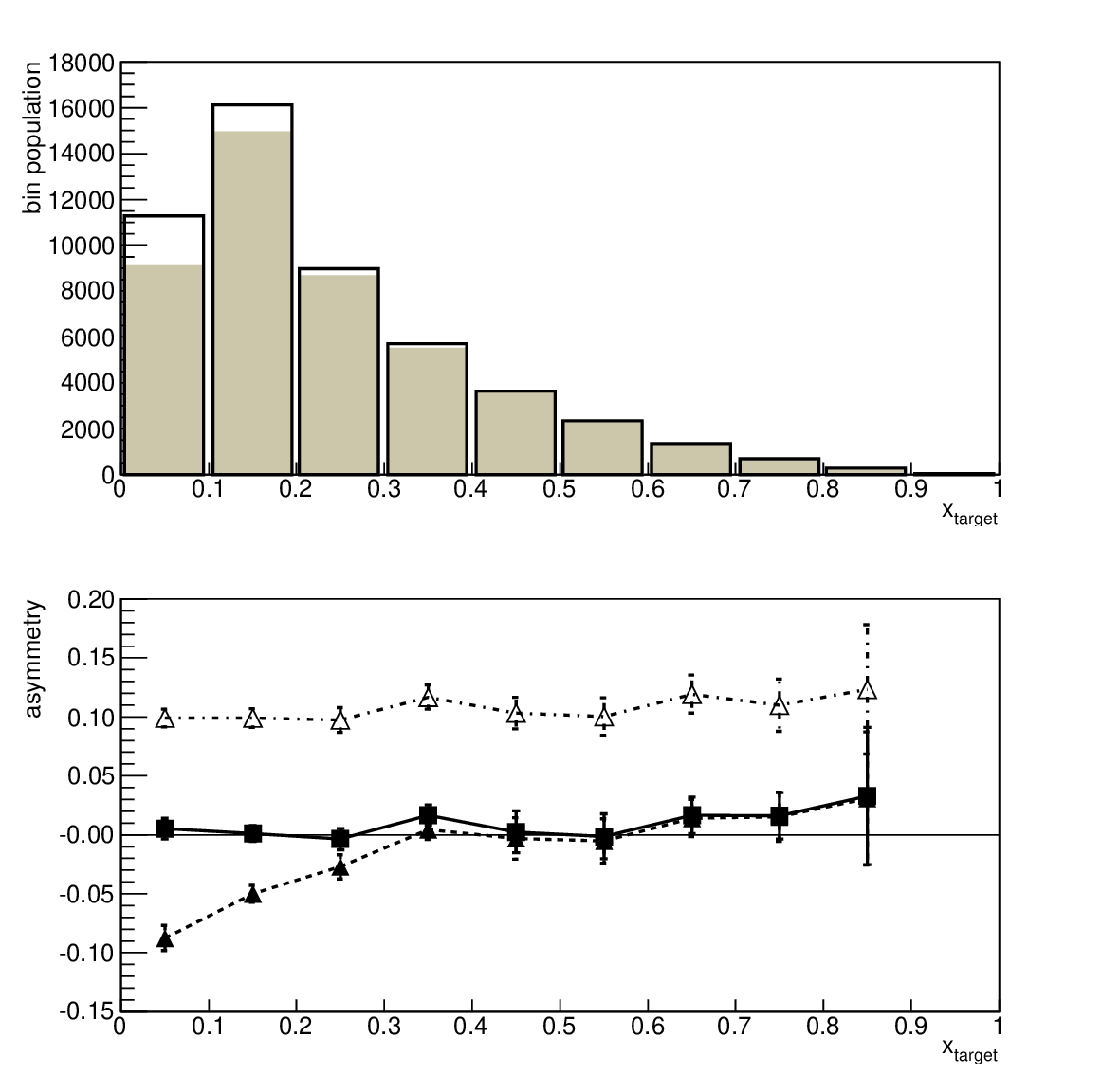}
\caption{
Simulated experiment in high-energy configuration with Method 1. 
In the upper plot the bin population is shown for each $x$ bin, without 
angular cutoff (black-line histogram) and with the cutoff 
$\theta_{lab\pm}$ $>$ 0.5$^\circ$ (shadowed histogram). 
In the lower plot, the extracted 
$cos(2\phi)$ asymmetry is shown. 
Continuous line: no physical 
asymmetry and no angular cutoff. 
Dot-dashed line: 
physical asymmetry 
with full phase space value 10 \%, 
no angular cuts. Dashed line: no physical asymmetry, 
cutoff $\theta_{lab\pm}$ $>$ 0.5$^\circ$ cut.
The error bars represent individual experiment errors. 
\label{fig:Xhnrg}}
\end{figure}

Here we adopt Method 1 of section 2.5 
to simulate the 
outcome of an experiment in PANDA configuration (fig.\ref{fig:Xpanda}) 
and high-energy configuration (fig. \ref{fig:Xhnrg}). 
The events are divided into 10 bins of the target longitudinal fraction.  
In each bin the events are integrated over the longitudinal 
fraction of the projectile and all the other variables. 

In both figures the event distributions 
with and without a forward cutoff are reported in the upper panel, 
the corresponding asymmetries in the lower panel. 
The curve close to zero values is the 
``reference'' one, i.e. the corresponding simulations were performed 
with neither physical asymmetries nor angular cutoffs. Within 
the reported error bars, it coincides with zero everywhere. 

In the PANDA configuration case (fig. \ref{fig:Xpanda}) 
the positive-valued curve is a physical asymmetry with full-phase-space 
asymmetry 6.5 \%, simulated in absence of 
any angular cutoff, 
according to the $x$-independent  
$\nu$-distribution with the same $Q_T$-dependence as in 
\cite{BRMCa}. The 
negative-valued curve is the fake asymmetry due to a forward angular cutoff 
5$^o$. 

In the high-energy configuration case (fig. \ref{fig:Xhnrg}) 
the positive-valued curve is a physical asymmetry with full-phase-space 
asymmetry 10 \%, simulated in absence of any angular cutoff,  
according to the $x$-independent $\nu$-distribution with the same 
$Q_T$-dependence as in \cite{BRMCa}. 
The negative-valued curve is the fake asymmetry due to a forward 
angular cutoff 0.5$^o$. 

\subsection{Dependence of the fake asymmetry on the size of the angular cutoff}

Here we use Method 1 
to explore the dependence of the fake asymmetry on the size of the 
angular cutoff. No physical asymmetry is present. The results refer 
to the full phase space integrated asymmetry. 

\begin{figure}[ht]
\centering
\includegraphics[width=9cm]{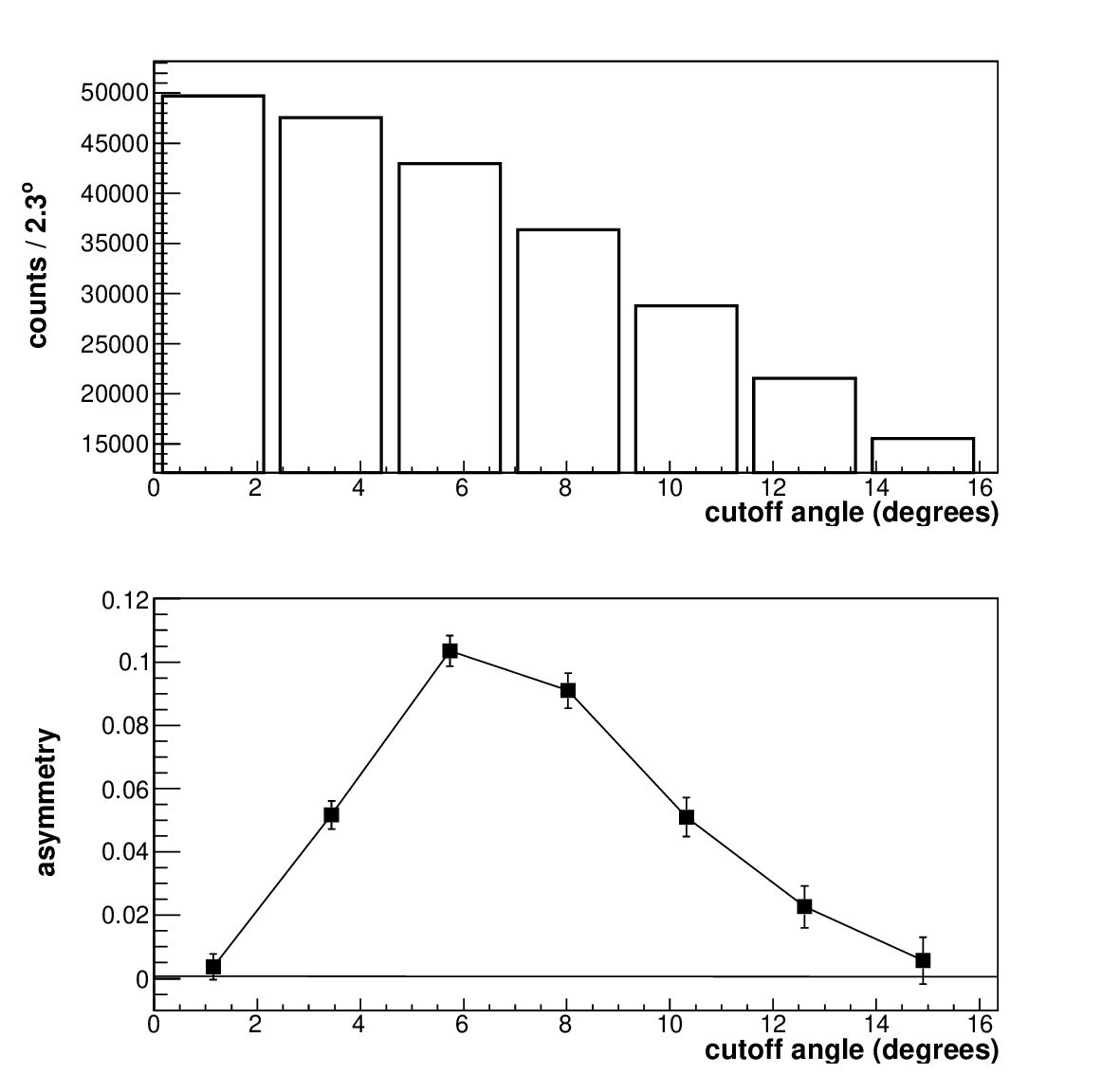}
\caption{
Dependence of the fake asymmetry on the angular cutoff  
in PANDA configuration. 
Each point corresponds to a different forward cutoff angle. 
Upper panel: average  
event population after the forward angle cutoff has been 
applied. 
Lower panel: average overall asymmetry. 
Asymmetries and event numbers are extracted via Method 1. The 
error bars represent errors on the theoretical estimates.
\label{fig:Xcutoff1}}
\end{figure}

Fig. \ref{fig:Xcutoff1} refers to PANDA configuration. The
applied cutoffs range from 0$^o$ to 15$^o$. 

\begin{figure}[ht]
\centering
\includegraphics[width=9cm]{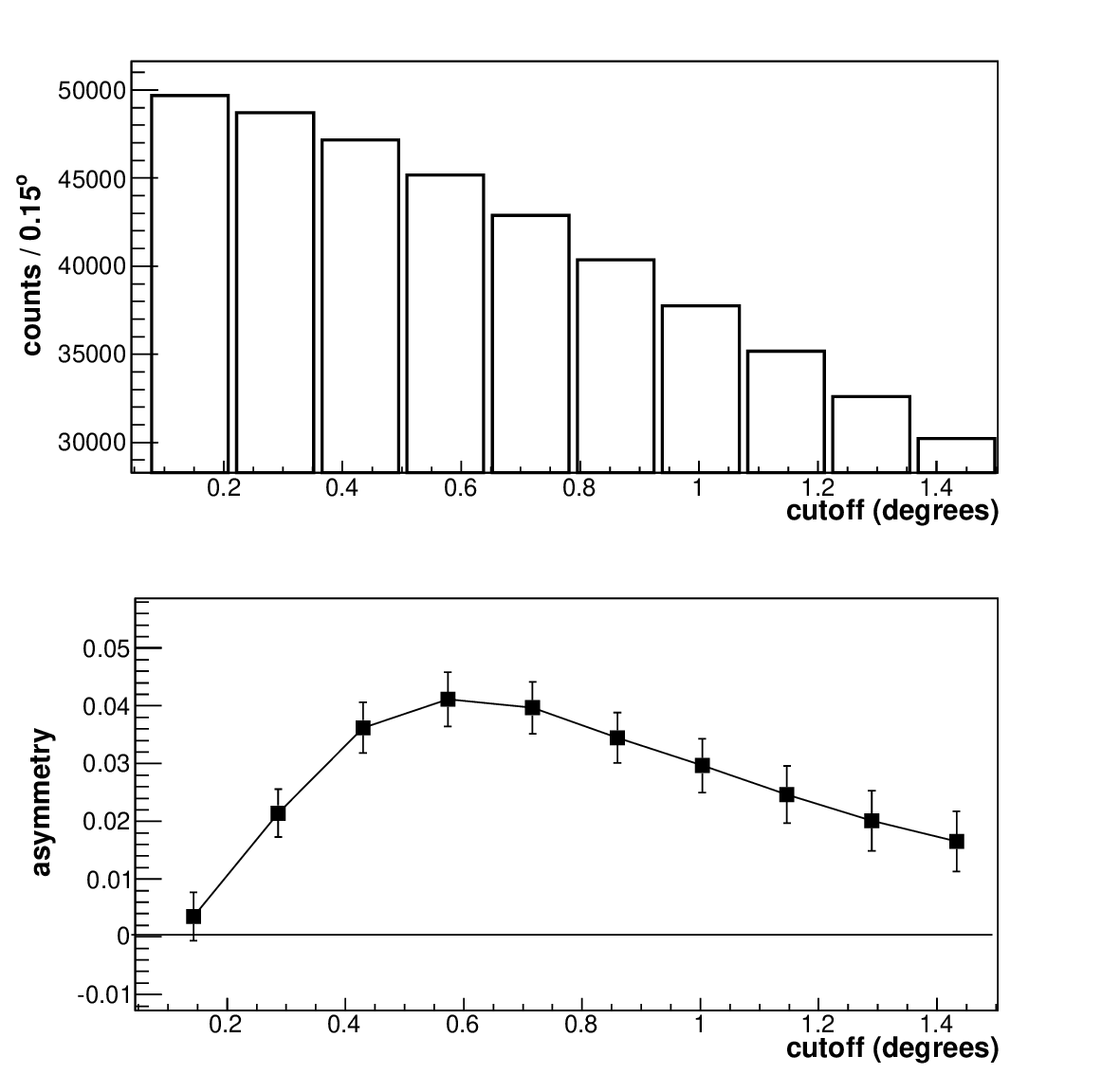}
\caption{
Dependence of the false asymmetry on the angular cutoff,  
in high-energy configuration with the mass range restricted 
to 4.5 $<$ $Q$ $<$ 5.5 GeV. 
Each point corresponds to a different forward cutoff angle. 
Upper panel: average  
event population after the forward angle cutoff has been 
applied. 
Lower panel: average overall asymmetry. 
Asymmetries and event numbers are extracted via Method 1. The 
error bars represent errors on the theoretical estimates.
\label{fig:Xcutoff2}}
\end{figure}

Fig. \ref{fig:Xcutoff2} refers to the high energy configuration 
with restricted mass range 4.5 GeV $<$ $Q$ $<$ 5.5 GeV. The applied 
cutoffs range from zero to 1.5$^o$. 

The biggest asymmetry is produced by angular cutoffs 
6$^o$ and 0.6$^o$ for the two configurations respectively. These cutoff 
values will be used in the following to examine the dependence of the effect 
on transverse momentum, mass, and other cutoffs. 


\subsection{Dependence of the fake asymmetry on mass and transverse momentum}  

These dependencies are considered in the high-energy configuration 
scheme, where an experiment may collect events in a broad range of $Q$  
and $Q_T$. No physical asymmetry is included. 

\begin{figure}[ht]
\centering
\includegraphics[width=9cm]{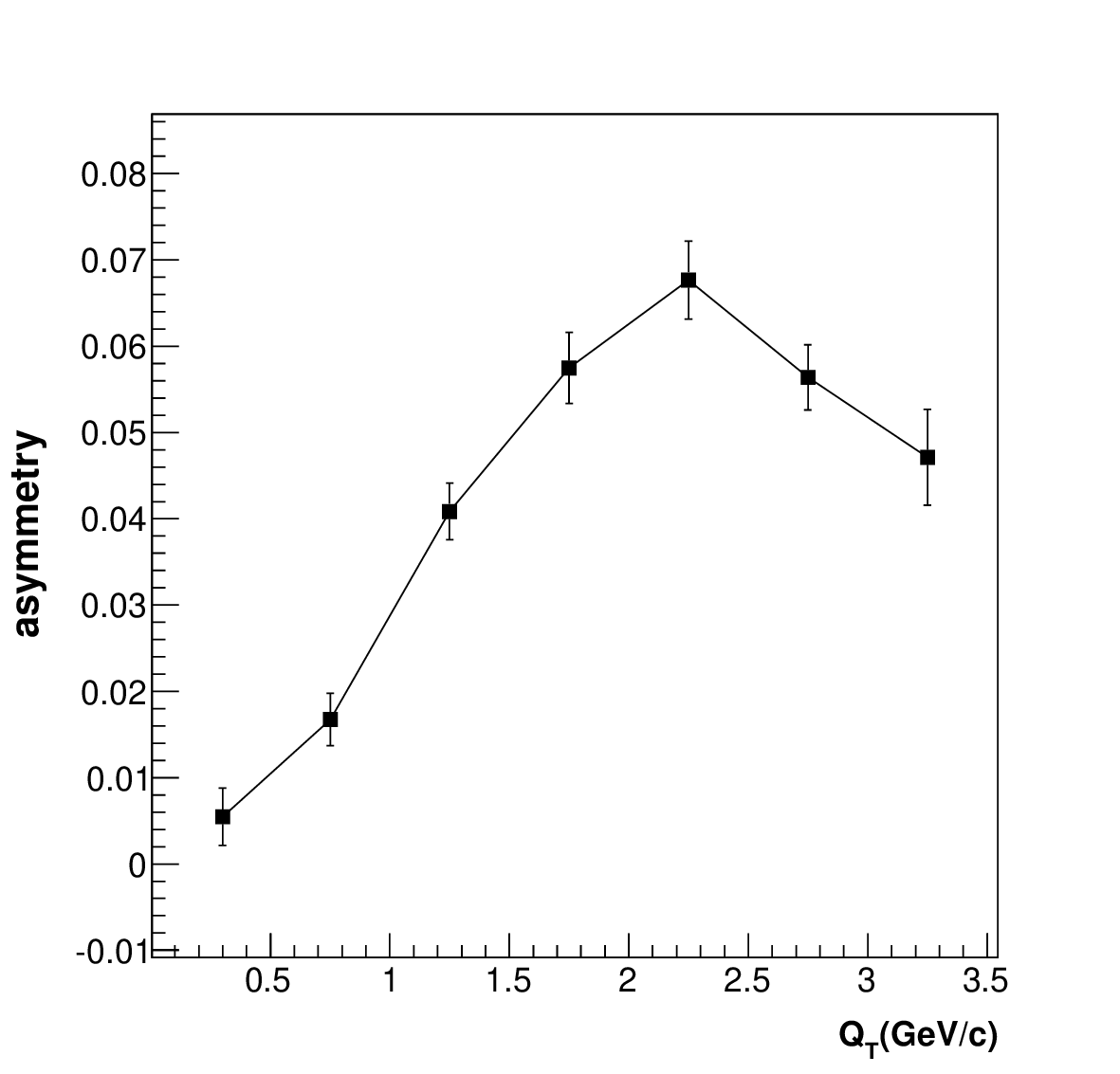}
\caption{
$Q_T-$dependence of the false asymmetry  
in high-energy configuration, with the mass range restricted to
4.5 $<$ $Q$ $<$ 5.5 GeV. 
For all the points the angular cutoff is $0.6^o$. 
These values have been extracted via Method 1. 
The error bars represent errors on the theoretical estimates. 
\label{fig:Xqt}}
\end{figure}

Fig.\ref{fig:Xqt} has been calculated in the restricted mass range 
4.5 GeV $<$ $Q$ $<$ 5.5 GeV, with Method 1. 
The cutoff angle is $0.6^o$, that according to fig.\ref{fig:Xcutoff2} 
produces the maximum fake asymmetry 
effect in the high energy configuration. 

\begin{figure}[ht]
\centering
\includegraphics[width=9cm]{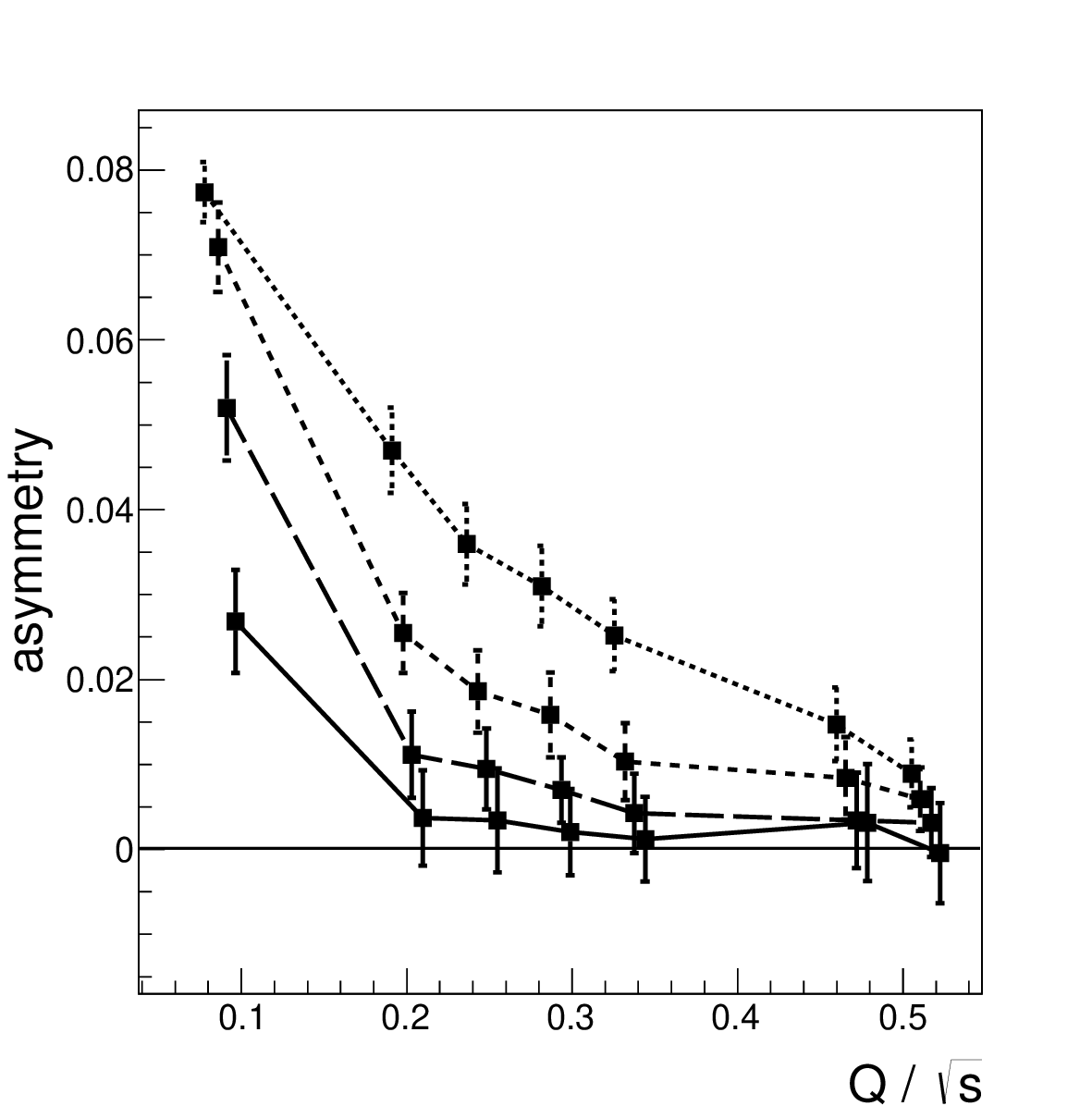}
\caption{
Dependence of the fake asymmetry on the dilepton mass $Q$
for several cutoffs in the laboratory angle, in high-energy 
configuration. 
Each curve corresponds to a different 
angular cutoff in the laboratory. 
Dotted curve: $0.6^o$. Short-dashed curve: $1.2^o$. 
Long-dashed curve: $1.8^o$. Continuous curve: $2.4^o$. 
For each curve, the presented points correspond 
to $Q-$ranges 1.5-2.5 GeV, 4-5 GeV, 5-6 GeV, 6-7 GeV, 7-8 GeV, 10-11 GeV, 
11-12 GeV. 
For each mass range, the corresponding four points indicate the 
central value of the range, although we have slightly shifted them 
to avoid graphical overlap of the error bars. These values have been 
extracted via Method 1, independently applied to each $Q$-interval. 
The error bars 
represent errors on the theoretical estimates. 
\label{fig:Xtau}}
\end{figure}

Fig. \ref{fig:Xtau} shows the dependence of the fake asymmetry effect on 
the virtual photon mass. 
Each mass bin has been treated as an independent experiment to 
which Method 1 has been applied, without sub-binning 
in any variable. The charmonium and bottonium regions have 
been excluded. Each curve in the figure corresponds to one 
value of forward cutoff: $0.6^o$, $1.2^o$, $1.8^o$, $2.4^o$. 

\subsection{Interaction with other cuts affecting the forward region}

Typically in Drell-Yan experiments additional cutoffs  
in $x_F$ $=$ $x_1-x_2$ 
and in 
$|cos(\theta)|$ are applied to the data. 
A cut on these parameters may remove lepton tracks from the forward 
region. In order to study its effect we have repeated the 
previous simulations 
imposing additional cutoffs on $|x_F|$ and in $|cos(\theta)|$.

The usual cut\cite{NA3a,NA3b,NA10a,NA10b,Conway89}  $|x_F|$ $<$ 0.9     
does not seem to introduce 
qualitative differences on the fake asymmetry. Stricter cuts on 
$|x_F|$ are not customary. 

The effect of a cut on $|cos(\theta)|$ is more interesting, 
since limits $|cos(\theta)|$ $<$ 0.8$-$0.9 are 
the rule in Drell-Yan experiments to remove effects due to the 
rescattering in the nuclear target (see e.g. the discussion on the 
NA3 data analysis in \cite{NA3a}). 

\begin{figure}[ht]
\centering
\includegraphics[width=9cm]{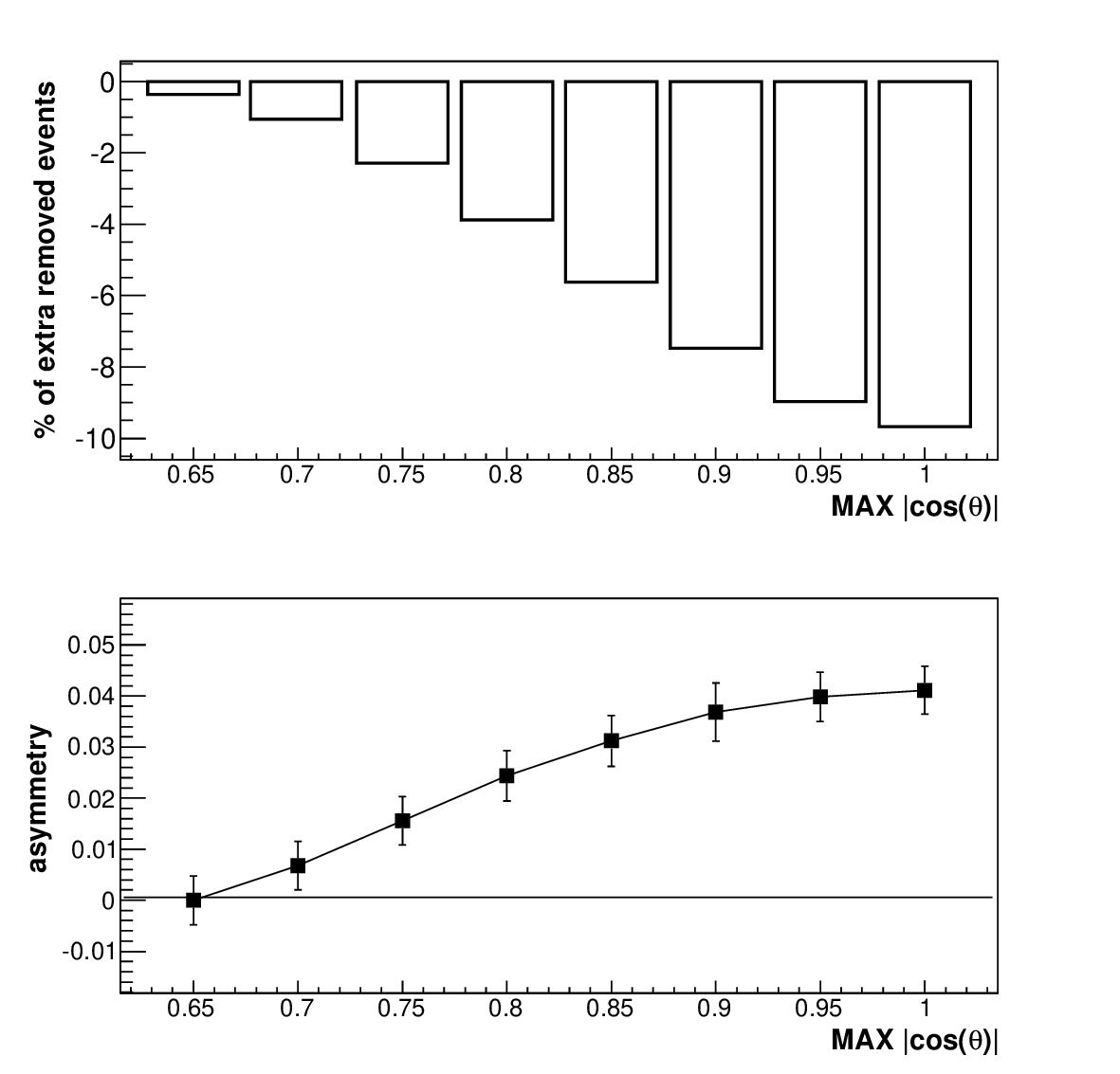}
\caption{
Effect of an additional cutoff on the Collins-Soper polar 
angle $\theta$ in high-energy configuration. 
A fixed cutoff $\theta_{Lab\pm}$ $>$ 0.6$^o$ is present in all  
the points. A further cutoff is applied on the Collins-Soper 
angle $\theta$: $|cos(\theta)|$ $<$ $Max|cos(\theta)|$. Each point 
corresponds to a different value of $Max|cos(\theta)|$. 
Upper panel: 
the percentage of the events removed by the laboratory cutoff  
after the cutoff on $|cos(\theta)|$ had been already 
applied.  
Lower panel: the fake asymmetry. 
These points have been reconstructed via Method 1. The error 
bars represent errors on the theoretical estimates. 
\label{fig:Xtheta}}
\end{figure}

We show the combined effect of a laboratory forward cutoff and of 
a Collins-Soper polar cutoff in fig. \ref{fig:Xtheta}. We work in 
the high-energy configuration. 
Two cutoffs are applied to the generated events: 

a) A cutoff on the Collins-Soper angle $\theta$: 

\noindent
$|cos(\theta)|$ $<$ $Max|cos(\theta)|$. 
Each point 
fig.\ref{fig:Xtheta} corresponds to a different value of 
$Max|cos(\theta)|$. 

b) A fixed cutoff $\theta_{Lab\pm}$ $>$ 0.6$^o$. This regards 
all the points in fig.\ref{fig:Xtheta}. 

For each of the eight points reported in the figure an independent 
simulation according to method 1 
is organized in the following way: 
The simulation generates exactly 
6x50,000 events $after$ applying the cut (a), but 
$before$ imposing the cut (b). At this point, the 
application of cut (b) removes some extra events. 
The percentage of the extra events removed by cut (b) 
is reported in the upper panel of 
fig.\ref{fig:Xtheta}. Clearly, this percentage is large when no 
cut on the 
Collins-Soper polar angle has been applied. When a strong cut of this 
kind has been applied,  
the laboratory cut acts on events that have already been 
partially removed. 

From 
the upper panel of fig.\ref{fig:Xtheta} we see that after applying 
the cutoff 
$|cos(\theta)|$ $<$ 0.7 (equivalent to $50^o$ $<$ $\theta$ $<$ 
$130^o$) the laboratory cutoff $\theta_{Lab\pm}$ $>$ 0.6$^o$ removes 
almost no events. This has the obvious consequence, visible in 
the lower panel, that the fake asymmetry reduces to zero in 
these conditions. 
 
\subsection{Interplay between true and fake asymmetries in presence of 
angular cutoffs} 

Another relevant point is the effect of an angular cutoff on a set 
of events where a true, physical azimuthal asymmetry is present. We 
have considered two possibilities: 

(i) a set of data where a physical $cos(2\phi)$ asymmetry is present, 

(ii) a set of data where a physical $cos(\phi)$ asymmetry is present. 

\noindent
Even the latter modifies  
the effect of the angular cutoff in producing an apparent 
$cos(2\phi)$ asymmetry\footnote{We 
have not considered the effects of 
the angular cutoffs on the $cos(\phi)$ asymmetry itself because 
this is out of the main scope of this work.
}. 

\begin{figure}[ht]
\centering
\includegraphics[width=9cm]{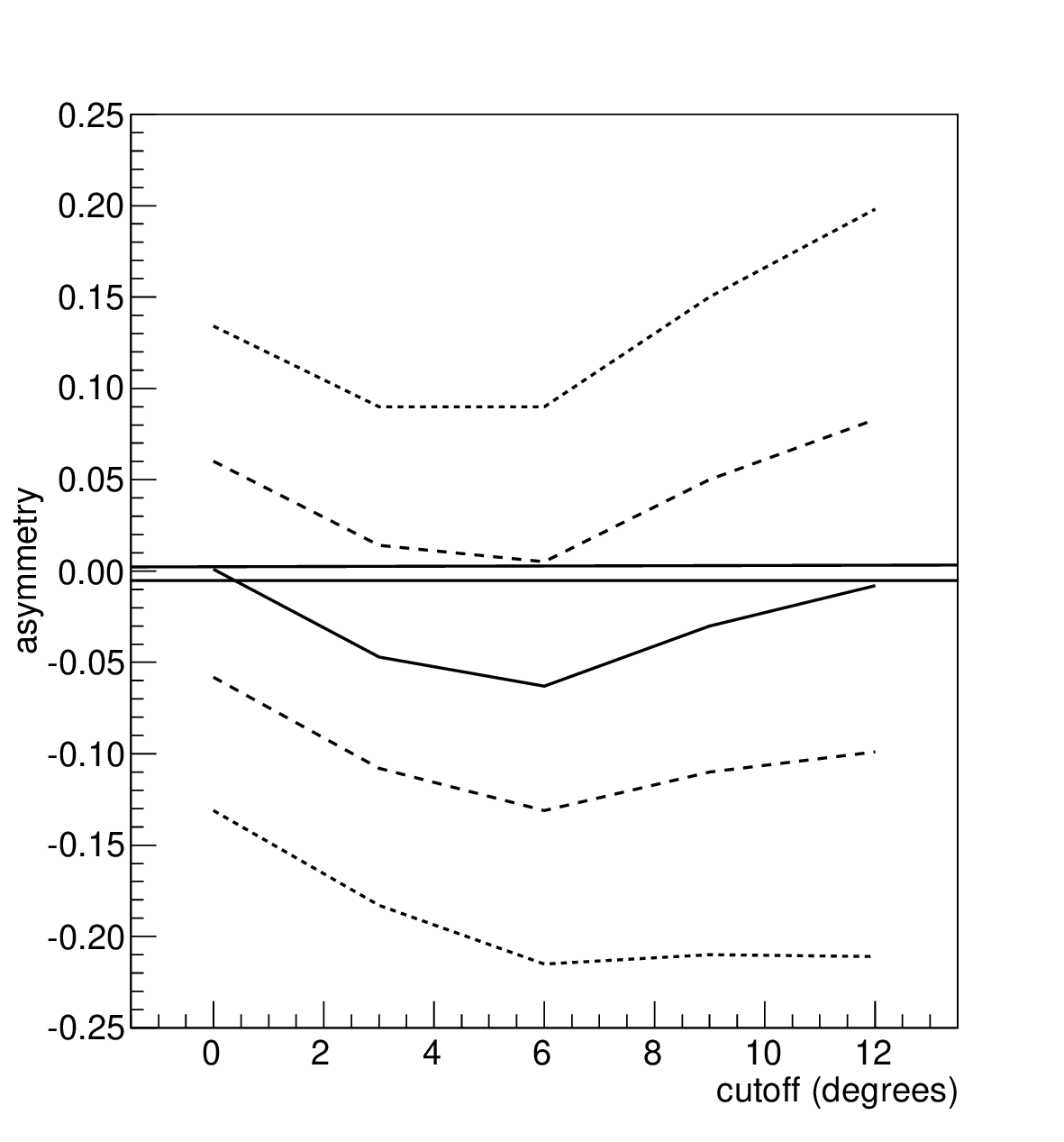}
\caption{
Dependence 
of the overall (fake plus real) $cos(2\phi)-$asymmetry 
on the cutoff angle, for different values of $A_{true-2\phi}$, that is the real  
asymmetry integrated over all the phase space, in PANDA configuration. 
Continuous 
line: $A_{true-2\phi}$ $=$ 0. Long-dashed lines: $A_{true-2\phi}$ $=$ 
$\pm$0.065. Short-dashed lines: $A_{true-2\phi}$ $=$ $\pm$ 0.135. 
The continuous almost horizontal lines show the $\pm\sigma$ error band. 
These points have been extracted via Method 2. 
\label{fig:a2p2p}}
\end{figure}

\begin{figure}[ht]
\centering
\includegraphics[width=9cm]{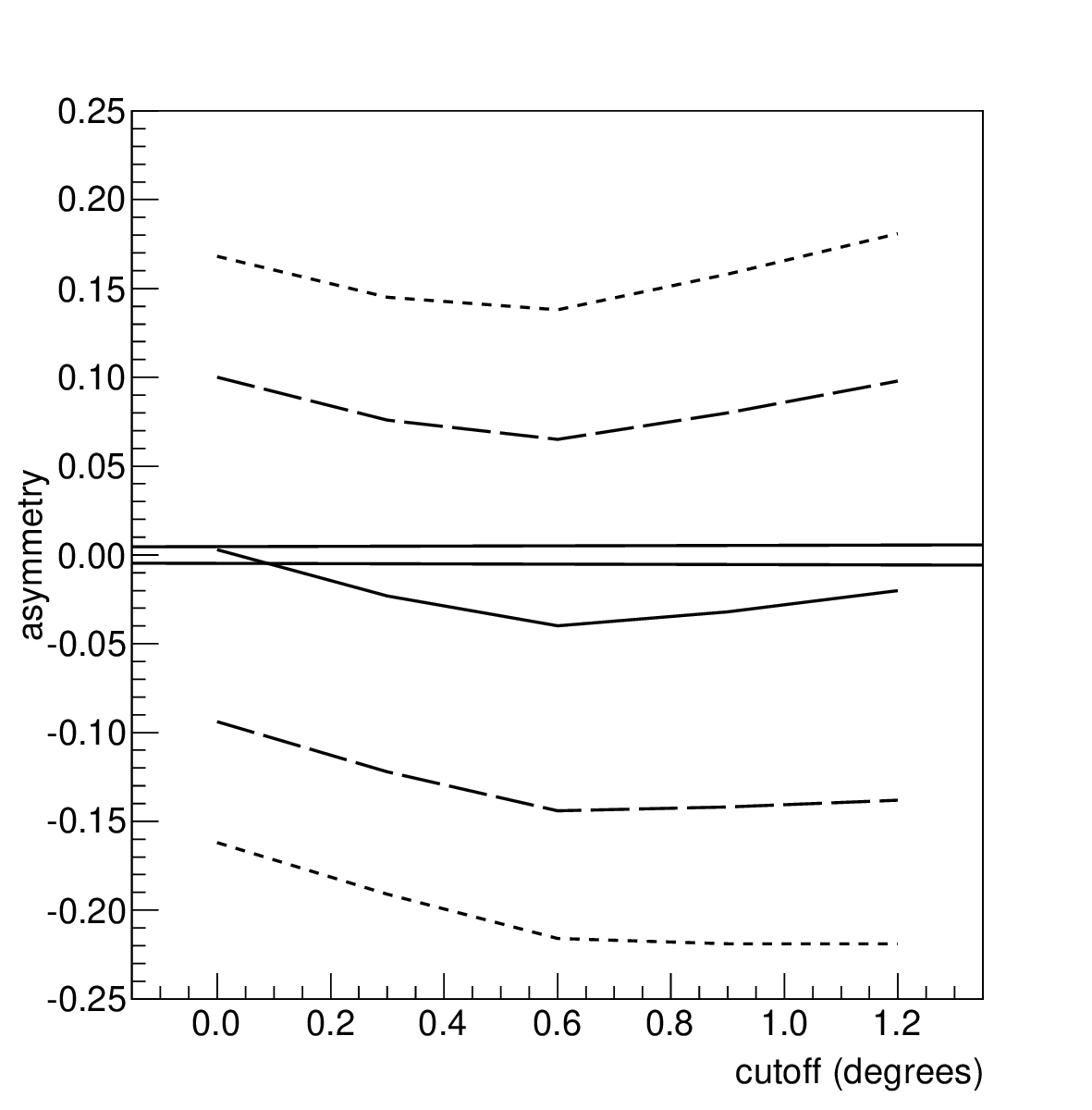}
\caption{
Dependence 
of the overall (fake plus real) $cos(2\phi)-$asymmetry 
on the cutoff angle, for different values of $A_{true-2\phi}$, that is the real  
asymmetry integrated over all the phase space, in high energy 
configuration (see text). Continuous 
line: $A_{true-2\phi}$ $=$ 0. Long-dashed lines: $A_{true-2\phi}$ $=$ 
$\pm$0.065. Short-dashed lines: $A_{true-2\phi}$ $=$ $\pm$ 0.135. 
The continuous almost horizontal lines show the $\pm\sigma$ error band. 
These points have been extracted via Method 2. 
\label{fig:b2p2p}}
\end{figure}

\begin{figure}[ht]
\centering
\includegraphics[width=9cm]{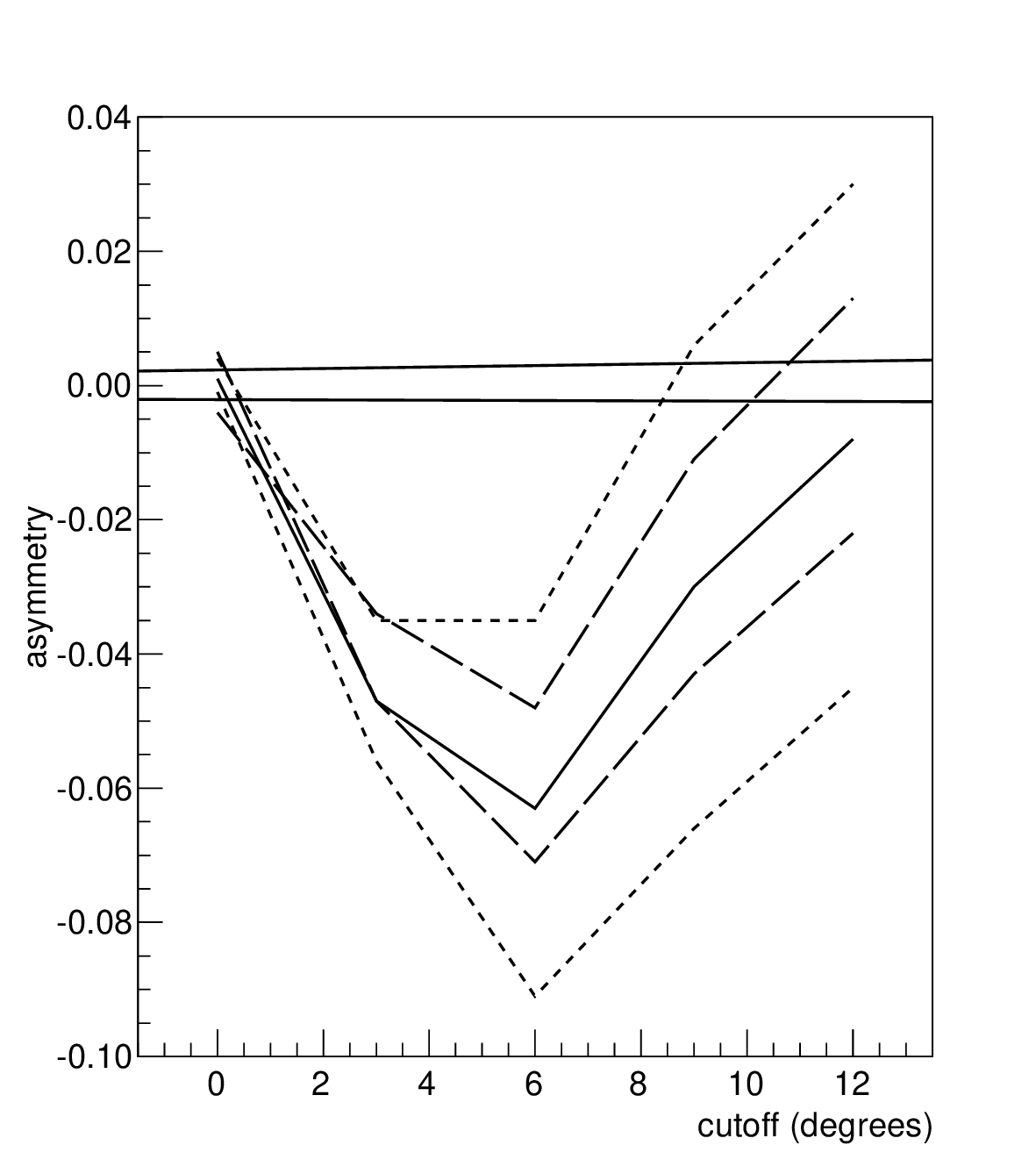}
\caption{
Dependence 
of the overall (fake plus real) $cos(2\phi)-$asymmetry 
on the cutoff angle, for different values of $A_{true-\phi}$, that is the real  
$sin(2\theta)cos(\phi)-$asymmetry 
integrated over all the phase space, in PANDA configuration. 
Continuous 
line: $A_{true-\phi}$ $=$ 0. Long-dashed lines: $A_{true-\phi}$ $=$ 
$\pm$0.1. Short-dashed lines: $A_{true-\phi}$ $=$ $\pm$ 0.16. 
Going from negative to positive $A_{true-\phi}$, the reported curves 
become more negative. 
The continuous almost horizontal lines show the $\pm\sigma$ error band. 
These points have been extracted via Method 2. 
\label{fig:a1p2p}}
\end{figure}

\begin{figure}[ht]
\centering
\includegraphics[width=9cm]{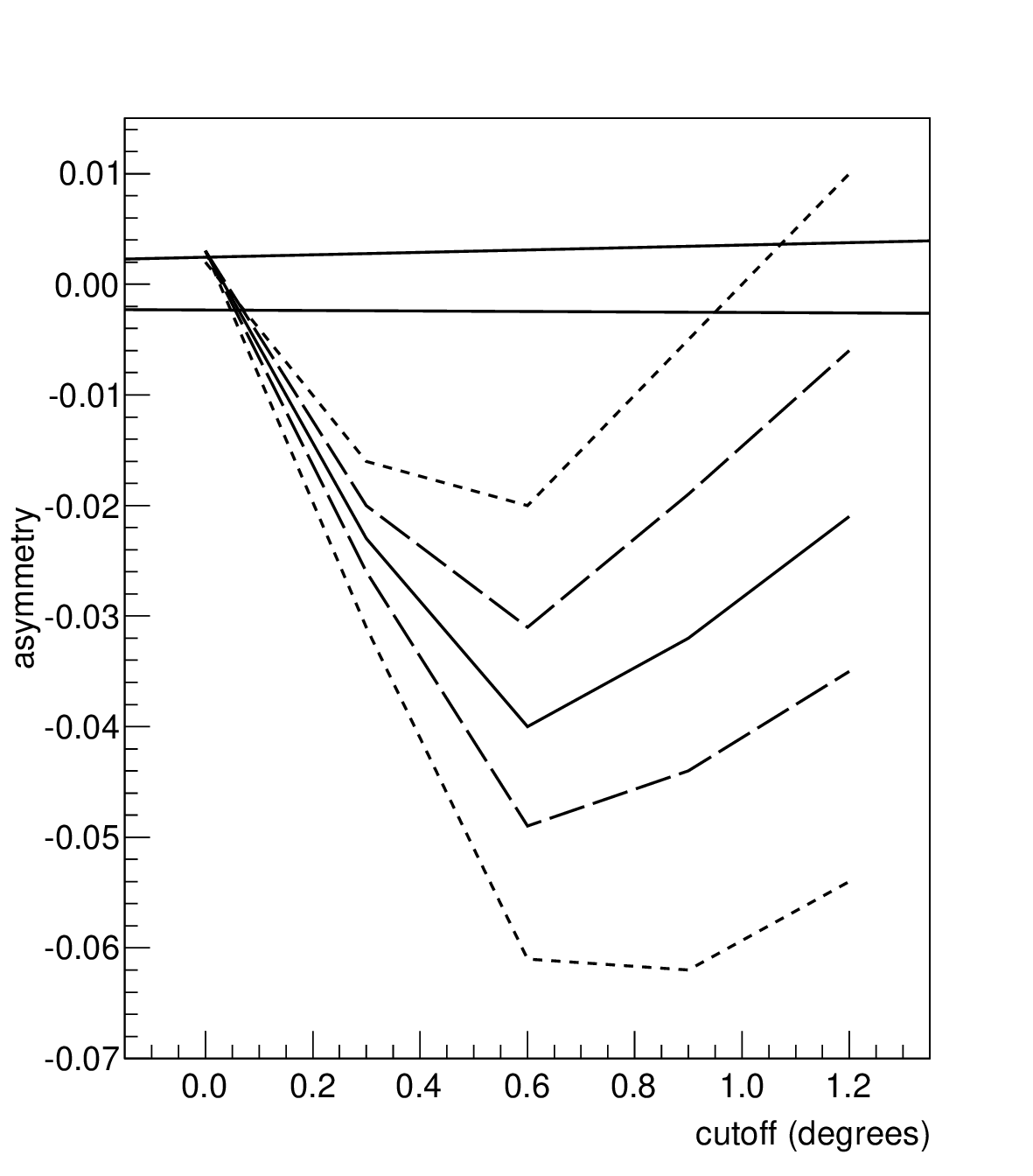}
\caption{
Dependence 
of the overall (fake plus real) $cos(2\phi)-$asymmetry 
on the cutoff angle, for different values of $A_{true-\phi}$, that is the real  
$sin(2\theta)cos(\phi)-$asymmetry 
integrated over all the phase space, in high energy configuration. 
Continuous 
line: $A_{true-\phi}$ $=$ 0. Long-dashed lines: $A_{true-\phi}$ $=$ 
$\pm$0.1. Short-dashed lines: $A_{true-\phi}$ $=$ $\pm$ 0.22. 
Going from negative to positive $A_{true-\phi}$, the reported curves 
become more negative. 
The continuous almost horizontal lines show the $\pm\sigma$ error band.  
These points have been extracted via Method 2. 
\label{fig:b1p2p}}
\end{figure}

The results discussed in this section are presented in figures 
\ref{fig:a2p2p}, \ref{fig:b2p2p}, \ref{fig:a1p2p}, and \ref{fig:b1p2p}.  

They have been calculated with Method 2.  
The associated error on the estimated asymmetries is $1/\sqrt{N}$ 
and ranges from 0.002 to 0.003. In each figure of this section 
it is represented by a narrow band on the x-axis. 
We notice (see fig.\ref{fig:a1p2p}) that this error is 
smaller than 
the fluctuation between the values of the $cos(2\phi)$ asymmetry in 
absence of angular cutoffs but in presence of different 
$cos(\phi)$ asymmetry. Since this fluctuation is pure 
error, it signals that in these simulations 
the numeric error is as relevant as the statistic one, and an overall 
error of size 0.005 may be estimated. 

We consider five different 
values of the physical $cos(2\phi)$ and $cos(\phi)$ asymmetries. 
Integrated over all the phase space, these are: 

\begin{itemize}
\item $cos(2\phi)$ asymmetry in PANDA and high-energy configuration: 
0, $\pm$ 6.5 \%, $\pm$ 13.5 \%. 

\item $cos(\phi)$ asymmetry in PANDA configuration: 
0, $\pm$ 10 \%, $\pm$ 16 \%. 

\item $cos(\phi)$ asymmetry in high-energy configuration: 
0, $\pm$ 10 \%, $\pm$ 22 \%. 
\end{itemize}

Each line in the figures \ref{fig:a2p2p} to \ref{fig:b1p2p} is 
associated to one of these values of physical asymmetry. A line shows 
how the  
measured $cos(2\phi)$ asymmetry changes at increasing angular 
cutoff.

The combined effects of a physical $cos(2\phi)$ asymmetry and of 
the angular cutoffs are reported in fig.\ref{fig:a2p2p} (PANDA 
configuration) and fig.\ref{fig:b2p2p} (high-energy configuration). 
These confirm the findings of figures \ref{fig:ad1},
\ref{fig:ad2}, \ref{fig:Xpanda}, and \ref{fig:Xhnrg}, i.e. 
that a forward cutoff causes a fake contribution 
of negative sign to the measured asymmetry. Therefore, 
the angular cutoffs may 
enhance or decrease the absolute value of the 
measured asymmetry, depending on the sign of the true underlying 
asymmetry. 

The joint effect of a nonzero physical $cos(\phi)$ asymmetry and of the 
angular cutoffs on the measured $cos(2\phi)$ asymmetry, in absence 
of a physical $cos(2\phi)$ asymmetry, is reported in 
fig.\ref{fig:a1p2p} (PANDA configuration) 
and fig.\ref{fig:b1p2p} (high-energy configuration). 

As we see, for a given angular cutoff 
there is $some$ dependence of 
the fake $cos(2\phi)$ asymmetry on the physical $cos(\phi)$ asymmetry.  
This dependence is however weak in the band $\pm$ 
10 \% of the physical asymmetry. If we compare 
figures \ref{fig:a2p2p} and \ref{fig:a1p2p} and observe the 
difference between the $\pm$ 10 \% curves at 6$^o$ where 
the cutoff-related distortion is maximum, this difference 
is over 12 \% in fig. \ref{fig:a2p2p}, but 
only 2.5 \% in fig. \ref{fig:a1p2p}. In the high energy 
configuration the difference between the two effects is even 
more marked (compare figures \ref{fig:b2p2p} and \ref{fig:b1p2p}
at 0.6$^o$). 

\section{Discussion} 

\subsection{Physical interpretation}

Speaking in very approximate terms, 
the Collins-Soper frame, seen from the laboratory frame, has the $z$-axis 
parallel to the beam, and the $x$-axis parallel to $\vec Q_T$. What 
really modifies the forward angle of the 
physical momenta when passing from the laboratory frame to the Collins-Soper 
frame is the boost, since in the latter  
the virtual photon is at rest. Consequently: 

\begin{itemize}

\item The angle $\theta$ is near $0^o$ when the positive muon 
is beam-parallel, near $180^o$ when the negative muon is beam-parallel. 
In these configurations the true $cos(2\phi)$ asymmetry is not 
present (see eq.\ref{eq:e2}), while the forward cutoff has its maximum 
effectiveness.  

\item $\theta$ is near $90^o$ when the momenta $q_+$ and $q_-$ of the 
muons in the laboratory have similar size, so that $\vec q_+-\vec q_-$ 
is transverse. In these kinematics the physical 
$cos(2\phi)$ asymmetry is strong. The 
forward laboratory cutoff does not seem to be effective: the upper 
panels of figures \ref{fig:ad1} and \ref{fig:ad2} show that for
$cos(\theta)$ $\approx$ $0^o$ event frequencies are not decreased 
by the forward cutoff. 

\item $\phi$ roughly coincides with the angle between the transverse 
component of $\vec q_+-\vec q_-$ and $\vec Q_T$ in the laboratory 
frame. In other words, with the angle between the lepton plane and 
the scattering plane (angle of rotation around the shared axis 
$\vec q$). 
A dominance of events where the transverse components of 
these two vectors are parallel or antiparallel means a positive 
$cos(2\phi)$ asymmetry. A dominance of events where they  
are orthogonal means a negative $cos(2\phi)$ asymmetry. 

\end{itemize}

The fact that an artificial asymmetry is present means that there 
are configurations with fixed photon momentum, fixed  
$|\vec q_+|$, $|\vec q_-|$ and fixed angle $\alpha_{+-}$ 
between the two leptons, where a rotation of the lepton plane on the 
$\vec q$-axis transforms an accepted 
event into a rejected event or does the opposite. In other words, a rotation 
of the lepton plane on the $\vec q$-axis 
sends a muon track into the dead cone, or takes a muon 
track out of it. 
In absence of a physical $cos(2\phi)$ asymmetry, all the rotations 
of the lepton plane on the photon axis produce equally likely 
events. If some of these rotations lead to rejected events, they 
are not equally likely anymore. 

In the situations examined in this work, the artificial asymmetry is 
always negative. Another way to describe the $cos(2\phi)$ asymmetry is 
that it is 
positive when one of the two muons is close to the beam axis and 
the other one is far, negative when their tracks have similar 
angles w.r.t. the beam. This explains the negative fake asymmetry 
suggesting that the events that are more frequently 
removed by a forward dead cone are those of the former kind, where one of 
the two tracks is dangerously close to the beam axis. 


If the qualitative interpretation is simple, quantitative predictions 
are difficult. The fake asymmetry effect depends on the interplay 
between 3 angles in the laboratory: 
the cutoff angle, the virtual photon angle 
$\approx$ $Q_T/Q_Z$, the angle $\alpha_{+-}$ 
between the two muons. 

If one of these three angles is zero, the fake asymmetry is zero: 
it becomes impossible to imagine configurations 
where a rotation of the lepton plane on the $\vec q$-axis   
transforms an accepted event into a rejected event. 
If one angle is not 
exactly zero, but it is anyway much smaller than the other two, 
we may imagine configurations where a rotation of the lepton plane 
on the $\vec q$-axis
transforms an accepted event into a rejected 
event, but these configurations are a small subset of the phase space. 

Something similar happens if one of the three angles is much 
larger than the other two. E.g., if the muon-muon angle is $10^o$ 
and the other 
two angles are $1^o$, the minimum angle between one muon and the 
beam axis is $9^o$ that is out of the dead cone. 

So, the fake asymmetry may be a statistically relevant effect at 
the condition that all these 3 angles have a similar magnitude. 
For a given cutoff angle, certain values of variables 
like $Q$, $Q_T$, or $x_{1,2}$ may statistically privilege 
values of the other two angles that are close each other, and 
close to the value of the cutoff angle. 


Let us consider the high energy configuration, where kinematics 
allows for some approximations. 

The virtual photon angle is $Q_T/Q_Z$. 
In the laboratory frame the energy of the virtual photon 
is mostly inherited by the beam parton, so 
$Q_Z$ $\approx$ $x_1E_{beam}$ $\approx$ $x_1 s/2M_{nucleon}$ 
$\approx$ $x_1 \cdot 270$ GeV.  
In absence of asymmetric cuts in $x_1$ and $x_2$, 
the most frequent value for $x_1$ is given by the two equations 
$x_1$ $\approx$ $x_2$, $x_1x_2$ $=$ $Q^2/s$, 
implying $x_1$ $\approx$ $\sqrt{Q^2/s}$ $\approx$ 0.23 for $Q$ $=$ 5 
GeV. A direct simulation confirms $<x_1>$ $=$ 0.23. 

This would lead to $Q_Z$ $\approx$ $270 \cdot 0.23$ 
$\approx$ 60 GeV. A direct simulation shows $Q_Z$ $\approx$ 
90 GeV. 
In the high-energy configuration $Q_T$ is constrained to the 
range 1-2.5 GeV/c and its simulated average value is 1.5 GeV/c, 
implying an angle $<Q_T>/<Q_Z>$ $\approx$ 1.5/90 that corresponds to 
$1^o$. 
The direct average of the ratio  
$<Q_T/Q_Z>$ in the simulated data gives 
1.35$^o$. In both cases 
the peak cutoff angle 0.6$^o$ extracted from fig.\ref{fig:Xcutoff2}
is about 1/2 of the virtual photon angle. 

In fig.\ref{fig:Xqt}, where the cutoff angle is fixed while 
$Q_T$ may freely assume any value in the range 0-3.5 GeV/c,
the largest fake asymmetry effect is for $Q_T$ $=$ 2-2.5 GeV/c. 
The simulation, restricted to $Q_T$ in the range 2-2.5 GeV/c, 
gives average photon angle 2$^o$, that is is more
than 3 times the cutoff angle 0.6$^o$.  

These distances between the cutoff angle and the photon 
angle may be justified observing in  
fig.\ref{fig:Xhnrg} that the fake asymmetry effect privileges 
small values of the target longitudinal fraction $x_2$, equivalent 
to large values of the projectile longitudinal fraction 
$x_1$ since $x_1x_2$ $\approx$ $Q^2/s$. This means that 
the relevant events present a virtual photon with $Q_Z$ larger 
than the average 90 GeV/c. This means that we have photon angles 
that are smaller than their global average, in those kinematic 
regions of $x_1$ and $x_2$ where the fake asymmetry effect 
is built. 

We know from the presented distributions that the fake asymmetry in the 
high-energy regime only regards 8 \% of the events at most.  
These events are likely to be fluctuations (as a small 
$x_2$ suggests) where the photon angle is especially small. 

It is more difficult to understand the role of 
the muon-muon angle $\alpha_{+-}$.  
Roughly, the average angle $<\alpha_{+-}>$ between the 
two leptons in the laboratory is $\propto$ $Q$, or to 
$\sqrt{x_2/x_1}$.  
Defining $y$ as the fraction of photon energy taken by the 
positive muon, we have 
$E_+$ $\equiv$ $yE_\gamma$, and $E_-$ $\equiv$ $(1-y)E_\gamma$, 
meaning $Q^2$ $\approx$ $y(1-y)E_\gamma^2 \alpha_{+-}^2$. 
Using 
$E_\gamma$ $\approx$ $Q_Z$ $\approx$ $x_1E_{beam}$ $\approx$ 
$x_1 s/2M_{nucleon}$ 
we get $x_2/x_1$ $\approx$ $\alpha_{+-}^2 y(1-y) s/4M_{nucleon}^2$.  

These approximations suggest that in fig.\ref{fig:Xtau} passing 
from $Q$ $=$ 2 GeV to $Q$ $=$ 12 GeV increases $<\alpha_{+-}>$ by six 
times, while in fig.\ref{fig:Xhnrg} increasing $x_{target}$ $\equiv$ 
$x_2$ means to increase $<\alpha_{+-}>$. Both 
operations cause the fake asymmetry effect to disappear. 

For symmetry, the most likely value of the average $\sqrt{y(1-y)}$ 
is 0.5, and $s/4M_{nucleon}^2$ $\approx$ 140.  
This gives $<\alpha_{+-}>$ $\approx$ $7^o \sqrt{x_2/x_1}$ suggesting 
a typical angle $7^o$. This may seem large, but a direct 
simulation in the mass range 4-8 GeV gave $<\alpha_{+-}>$ $\approx$ 
15$^o$ (average on the sphere, i.e. 
over $cos(\alpha_{+-}$). Restricting the mass range to 4-5 
GeV changes very little. 

What gives such a large average value 
to this angle are relatively rare decays in which one muon is 
backward-directed in the Collins-Soper 
frame, so to have small energy and a very large angle in the laboratory:
the average of nine 2$^o$ angles plus one 180$^o$ angle is 20$^o$. 

The peak of the distribution of $cos(\alpha_{+-})$ is at 3.5$^o$. 
Whether 
one prefers to attribute relevance to 3.5$^o$, to $7^o$ or to 
$15^o$ is a matter of taste, but all of them are much larger 
than the cutoff angle and the photon angle. 

However, reproducing the $cos(\alpha_{+-})$-distribution 
in the mass range 1.5-2.5 GeV shows that  
$<\alpha_{+-}>$ has more or less the same value as in the 4-8 GeV 
mass range, while the peak $\alpha_{+-}$ is now 1.5$^o$,  
not so far from the cutoff angle. The fact that at decreasing 
mass we have the same average angle, but a more forward peak in 
the distribution means that the population of the angles 
$<$ 1$^o$ increases at decreasing masses. 

Comparing this fact with 
the fast decrease of the fake asymmetry at increasing masses 
in fig.\ref{fig:Xtau}, we have good reasons to guess 
that the fake asymmetry is due to the small-angle part of the distribution 
of $\alpha_{+-}$, that is more populated at smaller $Q$. 
If we could extrapolate the 
Drell-Yan process to 
$Q$ $<$ 15 GeV\footnote{that 
is not possible since the vector mesons dominate the 
cross section, so another physics has to be 
taken into account. 
} 
we 
would find a maximum of the fake effect, and a decrease 
of the effect at lower masses, or equivalently lower 
$x_2$ for a given $x_1$. 

Summarizing this part, we expect to find a strong fake asymmetry 
effect when the cutoff angle, the virtual photon angle, and the 
lepton-lepton angle are similar. In the high energy configuration 
the average values of the virtual photon angle and of the 
lepton-lepton angle cannot realize this. 
Fluctuations from the averages of these two angles may combine 
so to do it, and this justifies the fact that the fake 
asymmetry is statistically a small effect, and that it is present  
at peculiar kinematics (e.g. small target $x$).  

\subsection{Perspectives} 

From fig. \ref{fig:Xtheta} it can be deduced that 
a strong $|cos(\theta)|$ 
cutoff assures a  clean asymmetry measurement even  without 
model-dependent Monte Carlo corrections. 
This fact is supported 
by the qualitative considerations at the beginning of the 
Discussion section. Removing broad angular regions implies the 
cost of a relevant reduction in the experiment statistics. However, 
in \cite{BRMCa} it was shown by direct simulations and assuming 
a nonzero physical asymmetry, that the optimal compromise between 
statistics and asymmetry dilution is reached with a rather 
strong cutoff in the Collins-Soper frame, excluding events 
out of the range $60^o$ $<$ $\theta$ $<$ $120^o$, i.e. 
requiring the cutoff $|cos(\theta)|$ $<$ 0.5. This cutoff is 
more restrictive than any cutoff reported in 
fig.\ref{fig:Xtheta}. The argument in \cite{BRMCa} 
was that the physical 
$cos(2\phi)$ asymmetry is proportional to the factor $sin^2(\theta)$ 
in eq.\ref{eq:e2}. As a consequence the removed events do not contribute  
to the asymmetry, but rather dilute it. 
 
The fake asymmetries induced by cuts on the acceptance in the forward region
feature a behavior that is similar in several respects 
to the one of the asymmetries measured in
the experiments previously cited:
both decrease at increasing $Q/\sqrt{s}$, both 
increase at increasing $Q_T$. 
The size of the fake asymmetry is smaller, but not 
by orders. 

With more than 
10,000 useful events, and in the same kinematic regime considered in 
this work (beam energies far below 1000 GeV), 
$\nu$ has 
been measured by the collaborations NA3\cite{NA3a,NA3b}, 
NA10\cite{NA10a,NA10b}, 
E615\cite{Conway89} in the decade 1980-90, leading to results that are 
similar in magnitude 
($\nu$ $\sim$ 0.1 at $Q_T$ $>$ 1 GeV/c) and present similar qualitative 
behaviors, within error bars. 
Although the related papers quoted here do not present a complete 
description of all the technical details of the data analysis, some 
figures are present showing a non-flat acceptance as a function of $\phi$ 
in the Collins-Soper frame. Therefore, we can assume that all the 
acceptance problems have been implicitly considered. 

Because of the limited phase space covered, in the NA10 experiment
the $|cos(\theta)|$ maximum was set to 0.5 or 0.6, depending on the 
run (see \cite{NA10a,NA10b}). This excluded the dangerous region of 
the forward angles.  
The corresponding fake asymmetry is negligible, as it may be 
deduced from fig. \ref{fig:Xtheta}. 
The experiments NA3\cite{NA3a,NA3b} and 
E615\cite{Conway89} put constraints on $cos(\theta)$ but not so severe. 
Their results are anyway similar to those from NA10. 

The procedure adopted by NA10 and NA3, according to the quoted papers, 
was expressing the theoretical cross section in terms of a set of 
parameters, convoluting this with the experiment known acceptance, 
generating events and comparing their distribution 
with the experimental outcome 
to select a reliable set of parameters including $\nu$. 
Although in some schemes 
the number of parameters may be as large as seven, it was remarked in 
\cite{NA3a} that the $\nu$-parameter is constrained by the 
distribution of the events coming from a specific phase space 
region ($\theta$ near 90$^o$) where the other parameters related with 
the angular distribution have little influence. So the search of the 
optimal value of the $\nu$-parameter was safer than the 
number of involved parameters could suggest. In the E615 
case\cite{Conway89} the technique was to integrate over two 
of the three angular parameters, 
while looking for a functional form $\nu(Q_T,x,...)$ for the third 
one that gave a satisfactory fit of the cross section behavior. 

In perspective, we should notice that to obtain a relative error 
$\pm$15\% on a single $\nu$ value summarizing the $Q$-range 4-4.5 GeV, 
the NA10 experiment had to concentrate 
40,000 events with $|cos(\theta)|$ $<$ 0.6 in that mass range. If 
these events were distributed in a finer $x-$binning, it would be 
difficult to reach a higher precision than $\pm$30\% in the most
populated bins. 
This suggests that very careful planning, and large event numbers, 
are necessary to allow the precision 
on the $\nu$-parameter to increase in magnitude w.r.t. the presently 
available values from the quoted experiments. 

\subsection{Conclusions}

We have shown that a forward dead cone in the laboratory 
frame seriously affects a measurement of a 
Lam-Tung-style lepton asymmetry. 
This could be a problem for experiments aiming at reducing 
the relative errors on the $\nu$-parameter  
to less than 20 \% in a broad $x$-range, in particular when 
the values of this parameter are needed as an input to 
extract Transversity from a single-spin Drell-Yan experiment. 

Our results show that the fake asymmetry is relevant for 
special values of the forward cutoff angle. Approximately, these 
correspond to situations where the cutoff angle, the polar 
angle and the angle betwen the two leptons have similar 
magnitudes in the laboratory. 

Fig.\ref{fig:Xtheta} 
suggests that restricting the data analysis to 
regions near $\theta$ $=$ 90$^o$ should be the safest way to 
obtain a result that is model-independent and precise. This 
agrees with the suggestion given in \cite{BRMCa},  
that including events from $\theta$-regions far from 
$90^o$ does not improve the quality of a measurement of 
the physical $cos(2\phi)$ asymmetry. 


\end{document}